\documentstyle[12pt,epsf]{article}
\newcommand{\logqmms}{l_{qm}}

\newcommand{\logqmums}{l_{q\mu}}

\newcommand{\logmsos}{L_{ms}}

\newcommand{\gsim}{\;\rlap{\lower 3.5 pt \hbox{$\mathchar \sim$}} \raise 1pt
 \hbox {$>$}\;}

\renewcommand{\thefootnote}{\fnsymbol{footnote}}

\begin{document}    

\title{\vskip-3cm{\baselineskip14pt
\centerline{\normalsize\hfill MPI/PhT/97--013}
\centerline{\normalsize\hfill TTP97--12\footnote{The 
  complete postscript file of this
  preprint, including figures, is available via anonymous ftp at
  www-ttp.physik.uni-karlsruhe.de (129.13.102.139) as /ttp97-12/ttp97-12.ps 
  or via www at http://www-ttp.physik.uni-karlsruhe.de/cgi-bin/preprints.}}
\centerline{\normalsize\hfill hep-ph/9704436}
\centerline{\normalsize\hfill April 1997}
}
\vskip1.5cm
Higgs Decay to Top Quarks at ${\cal O}(\alpha_s^2)$
}
\author{
 R.~Harlander$^{a}$ 
 and 
 M.~Steinhauser$^{b}$
}
\date{}
\maketitle

\begin{center}
$^a${\it Institut f\"ur Theoretische Teilchenphysik,
    Universit\"at Karlsruhe,\\ D-76128 Karlsruhe, Germany.\\ }
  \vspace{3mm}
$^b${\it Max-Planck-Institut f\"ur Physik,
    Werner-Heisenberg-Institut,\\ D-80805 Munich, Germany\\ }
\end{center}

\begin{abstract}
  \noindent Three-loop corrections to the scalar and pseudo-scalar
  current correlator are calculated.  By applying the large momentum
  expansion mass terms up to order $(m^2/q^2)^4$ are evaluated
  analytically. As an application ${\cal O}(\alpha_s^2)$ corrections to
  the decay of a scalar and pseudo-scalar Higgs boson into top quarks
  are considered. It is shown that for a Higgs mass not far above the
  $t\bar{t}$ threshold these higher order mass corrections are necessary
  to get reliable results.

\end{abstract}

\thispagestyle{empty}
\newpage
\setcounter{page}{1}


\renewcommand{\thefootnote}{\arabic{footnote}}
\setcounter{footnote}{0}

\section{Introduction and notation}

A crucial question in elementary particle physics is whether nature
makes use of the spontaneous symmetry breaking for generating the
masses or not.
In the minimal standard model (SM) this mechanism requires the 
existence of a scalar particle, the Higgs boson. 
Extensions of the SM, e.g. models with more than one Higgs doublet
or supersymmetric versions of the SM, predict also
pseudo-scalar Higgs bosons, $A$.

However, up to now there is no experimental evidence for such a particle
and the direct search at LEP via the process $e^+e^-\to f\bar{f}H$ rules
out the mass range $M_H\le 65.6$~GeV with a 95\% confidence level (CL)
\cite{Jan96}. Assuming the validity of the SM the precision data are
sensitive to the Higgs boson and a recent global fit yields
$M_H=149^{+148}_{-82}$~GeV together with a 95\% CL upper bound of
550~GeV \cite{Bou96} (in this context see also \cite{DegGamSir96}).
Theoretical arguments based on unitarity or the validity of perturbation
theory request an upper limit on the Higgs mass of about 1~TeV.

In this paper we consider the decay of a Higgs boson, scalar or
pseudo-scalar, with mass above 400~GeV, into a top-antitop quark pair.
Since the velocity of one of the quarks for this mass range is $v\gsim 0.5$,
threshold effects can safely be neglected and the decay rate can be
treated in a purely perturbative way. 
The full mass dependence for this process is
only known to ${\cal O}(\alpha_s)$ \cite{DreHik90}, while to order
$\alpha_s^2$ only mass terms up to $M_t^2/M_{H/A}^2$ are available
\cite{GorKatLarSur90,Sur94,CheKwi96,Sur942}, where $M_t$ is the 
top quark mass, and they show up to be 
quite sizeable \cite{Sur94}. 
${\cal O}(\alpha_s^3)$ corrections have recently been evaluated
in the massless case \cite{Che96}.

In the following, making use of a technique recently developed for the
automatic computation of mass corrections to the vector current
correlator \cite{CheHarKueSte96,CheHarKueSte97}, we will demonstrate
that the inclusion of mass terms up to order $(M_t^2/M_{H/A}^2)^4$ 
leads to reliable results at 
${\cal O}(\alpha_s^2)$ in the considered mass range.
We note that the results presented in this paper for the decay into top
quarks may be generalized to any fermion species.  However, even for
bottom quarks already the quadratic mass corrections are very small.

To fix the notation we define:
\begin{eqnarray}
  q^2\,\Pi^\delta(q^2)&=&i\int dx\,e^{iqx} 
  \langle 0|Tj^\delta(x)j^\delta(0)|0 \rangle,
  \,\,\,\mbox{with  }\,\,\, \delta\in\{s,p\}
\end{eqnarray}
where $\Pi^{s}(q^2)$ ($\Pi^{p}(q^2)$) represents the scalar (pseudo-scalar)
current correlator in momentum space.
The currents are given by
\begin{eqnarray}
j^s = \bar{\psi}\psi, &\qquad&
j^p = i \bar{\psi}\gamma_5 \psi.
\end{eqnarray}
It should be noted that as far as renormalization is concerned it is
convenient to consider the combination $mj^\delta$
in the on-shell scheme and $\bar{m}j^\delta$ in the $\overline{\mbox{MS}}$
scheme, where $m$ ($\bar{m}$) is the generic pole ($\overline{\mbox{MS}}$)
mass, in order to avoid
additional renormalization constants.  The physical observable
$R^\delta(s)$ is related to $\Pi^\delta(q^2)$ via the relation
\begin{eqnarray}
R^\delta (s) &=& 8\pi\,\mbox{Im}\,\Pi^\delta(q^2=s+i\epsilon).
\label{eqrtopisp}
\end{eqnarray}
The current correlator can be written as
\begin{eqnarray}
\Pi^\delta(q^2) &=& \Pi^{(0),\delta}(q^2) 
         + \frac{\alpha_s(\mu^2)}{\pi} C_F \Pi^{(1),\delta}(q^2)
         + \left(\frac{\alpha_s(\mu^2)}{\pi}\right)^2\Pi^{(2),\delta}(q^2)
         + \ldots\,\,,
\\
\Pi^{(2),\delta} &=&
                C_F^2       \Pi_A^{(2),\delta}
              + C_A C_F     \Pi_{\it NA}^{(2),\delta}
              + C_F T   n_l \Pi_l^{(2),\delta}
              + C_F T       \Pi_F^{(2),\delta}
              + C_F T       \Pi_S^{(2),\delta},
\label{eqpi2}
\end{eqnarray}
and similarly for $R^\delta(s)$.
The normalization in Eq.~(\ref{eqrtopisp}) guarantees that
$R^{(0),\delta}(s)\to 3$ for $s\to\infty$.
The colour factors ($C_F=(N_c^2-1)/(2N_c)$ and $C_A=N_c$)
correspond to the Casimir operators of the fundamental
and adjoint representations, respectively.
For the numerical evaluation we set $N_c=3$.
The trace normalization of the fundamental representation is $T=1/2$.
The number of light (massless) quark flavours is denoted by $n_l$.

In Eq.~(\ref{eqpi2}) $\Pi_A^{(2),\delta}$ is the abelian contribution
which also exists in QED, and $\Pi_{NA}^{(2),\delta}$ results from the
non-abelian structure specific for QCD. The contribution of diagrams
containing a second massless or massive quark loop is denoted by
$\Pi_l^{(2),\delta}$ and $\Pi_F^{(2),\delta}$, respectively.
$\Pi_S^{(2),\delta}$ represents the terms arising from
the double-triangle diagram
and is called the singlet contribution.  The case for the double-bubble
diagram where the inner quark mass is much heavier than the outer one is
not listed as it does not contribute in cases of physical
interest.

The outline of the paper is as follows:
In Section~\ref{seclmp} the method is briefly described and the
results for the correlator functions are given.
In Section~\ref{sechtt} mass effects for the decay of a scalar 
and pseudo-scalar Higgs particle into top quarks are presented 
and the results are discussed.


\section{\label{seclmp}Results for the current correlators}

In this section we keep the discussion general and consider 
the scalar and pseudo-scalar current correlators with a 
generic quark mass $m$. In Section~\ref{sechtt} the 
specification to the case $m=M_t$ is performed and numerical values are
given. 

Let us in a first step outline the main ideas of the large momentum
expansion and its realization by a computer\footnote{ For a more
  detailed discussion see Ref.~\cite{CheHarKueSte97}.}.  The large
momentum procedure \cite{Smi95} requires the identification of certain
subgraphs associated with the Feynman diagram to be calculated.  These
subgraphs then have to be expanded in their small dimensional quantities
(i.e.~all except the large momenta) and the resulting terms, being
products of tadpoles and massless integrals, have to be calculated. The
number of these terms increases rapidly with the number of loops. In our
case of a massive current correlator, the one- and two-loop diagrams
generate 17 terms altogether, so that they still can be treated by hand.
In the three-loop case, however, there are 19 topologies, contributing
266 terms, and their generation by hand is not feasible.  Therefore we
completely automated the large momentum procedure for massive two-point
functions and directly fed the generated terms to the FORM \cite{form}
packages MATAD and MINCER \cite{mincer}.  After calculating the single
terms the results of all diagrams were added giving a finite expression
after renormalization.

We are now prepared to present the result.
Let us first consider the scalar correlator.
The results for the different contributions read in the 
$\overline{\mbox{MS}}$ scheme ($\logqmms \equiv \ln(-q^2/\bar{m}^2), 
\logqmums \equiv \ln(-q^2/\mu^2))$:
\begin{eqnarray}
   \bar{\Pi}^{(0),s} &=& {3\over 16 \pi^2} \bigg\{
       4
          - 2\,\logqmums
       + {\bar{m}^2\over q^2}\, \bigg[
          - 16
          + 12\,\logqmums
          \bigg]
       + \left({\bar{m}^2\over q^2}\right)^{2} \, \bigg[
          - 18
          - 12\,\logqmms
          \bigg]
\nonumber\\&&\mbox{}
       + \left({\bar{m}^2\over q^2}\right)^{3} \, \bigg[
          {4\over 3}
          - 8\,\logqmms
          \bigg]
       + \left({\bar{m}^2\over q^2}\right)^{4} \, \bigg[
          7
          - 12\,\logqmms
          \bigg]
\bigg\}
 + \ldots\,\,,
\label{eqpi0s}
         \\[.4cm]
   \bar{\Pi}^{(1),s} &=& {3\over 16 \pi^2} \bigg\{
       {131\over 8}
          - 6\,\zeta_3
          - {17\over 2}\,\logqmums
          + {3\over 2}\,\logqmums^2
\nonumber\\&&\mbox{}
       + {\bar{m}^2\over q^2}\, \bigg[
          - 94
          + 36\,\zeta_3
          + 60\,\logqmums
          - 18\,\logqmums^2
          \bigg]
\nonumber\\&&\mbox{}
       + \left({\bar{m}^2\over q^2}\right)^{2} \, \bigg[
          - 44
          - 48\,\zeta_3
          - 69\,\logqmms
          - 18\,\logqmms^2
          + \left(63
              + 54\,\logqmms \right) \logqmums
          \bigg]
\nonumber\\&&\mbox{}
       + \left({\bar{m}^2\over q^2}\right)^{3} \, \bigg[
          - {461\over 9}
          - {956\over 9}\,\logqmms
          - {98\over 3}\,\logqmms^2
          + \left(- 20
              + 48\,\logqmms \right) \logqmums
          \bigg]
\nonumber\\&&\mbox{}
       + \left({\bar{m}^2\over q^2}\right)^{4} \, \bigg[
          {8263\over 144}
          - {881\over 6}\,\logqmms
          - {135\over 2}\,\logqmms^2
          + \left(- {141\over 2}
              + 90\,\logqmms \right) \logqmums
          \bigg]
\bigg\}
 + \ldots\,\,,
\label{eqpi1s}
         \\[.4cm]
   \bar{\Pi}^{(2),s}_{\it A} &=& {3\over 16 \pi^2} \bigg\{
       {1613\over 64}
          - 24\,\zeta_3
          + {9\over 4}\,\zeta_4
          + 15\,\zeta_5
          + \left(- {691\over 32}
              + {9\over 2}\,\zeta_3 \right) \logqmums
          + {105\over 16}\,\logqmums^2
          - {3\over 4}\,\logqmums^3
\nonumber\\&&\mbox{}
       + {\bar{m}^2\over q^2}\, \bigg[
          - {3471\over 16}
          + 220\,\zeta_3
          - 18\,\zeta_4
          - 130\,\zeta_5
\nonumber\\&&\mbox{\hspace{.2cm}}
          + \left({1911\over 8}
          - 72\,\zeta_3 \right) \logqmums
          - {369\over 4}\,\logqmums^2
          + 18\,\logqmums^3
          \bigg]
\nonumber\\&&\mbox{}
       + \left({\bar{m}^2\over q^2}\right)^{2} \, \bigg[
          - {4517\over 32}
          - 340\,\zeta_3
          + 72\,\zeta_4
          + 115\,\zeta_5
          - 12\,B_4
\nonumber\\&&\mbox{\hspace{.2cm}}
          + \left(- {2871\over 16}
              + 27\,\zeta_3 \right) \logqmms
          - {315\over 4}\,\logqmms^2
          - 18\,\logqmms^3
\nonumber\\&&\mbox{\hspace{0.4cm}}
          + \left({819\over 8}
              + 216\,\zeta_3
              + {1053\over 4}\,\logqmms
              + 81\,\logqmms^2 \right) \logqmums
          + \left(- {405\over 4}
              - {243\over 2}\,\logqmms \right) \logqmums^2
          \bigg]
\nonumber\\&&\mbox{}
       + \left({\bar{m}^2\over q^2}\right)^{3} \, \bigg[
          - {48895\over 1944}
          - 427\,\zeta_3
          + 48\,\zeta_4
          - 80\,\zeta_5
          - 8\,B_4
\nonumber\\&&\mbox{\hspace{0.4cm}}
          + \left(- {8938\over 27}
              - 116\,\zeta_3 \right) \logqmms
          - {1375\over 6}\,\logqmms^2
          - {1639\over 27}\,\logqmms^3
\nonumber\\&&\mbox{\hspace{0.4cm}}
          + \left({291\over 2}
              + {1636\over 3}\,\logqmms
              + 196\,\logqmms^2 \right) \logqmums
          + \left(96
              - 144\,\logqmms \right) \logqmums^2
          \bigg]
\nonumber\\&&\mbox{}
       + \left({\bar{m}^2\over q^2}\right)^{4} \, \bigg[
          - {6460859\over 31104}
          - 60\,\zeta_3
          + 72\,\zeta_4
          - 40\,\zeta_5
          - 12\,B_4
\nonumber\\&&\mbox{\hspace{0.4cm}}
          + \left(- {1242845\over 1296}
              - 173\,\zeta_3 \right) \logqmms
          - {546029\over 864}\,\logqmms^2
          - {32735\over 216}\,\logqmms^3
\nonumber\\&&\mbox{\hspace{0.4cm}}
          + \left(- {63305\over 96}
              + 910\,\logqmms
              + {2025\over 4}\,\logqmms^2 \right) \logqmums
          + \left({2655\over 8}
              - {675\over 2}\,\logqmms \right) \logqmums^2
          \bigg]
\bigg\}
 + \ldots\,\,,
\label{eqpi2as}
         \\[.4cm]
   \bar{\Pi}^{(2),s}_{\it NA} &=& {3\over 16 \pi^2} \bigg\{
       {14419\over 288}
          - {75\over 4}\,\zeta_3
          - {9\over 8}\,\zeta_4
          - {5\over 2}\,\zeta_5
          + \left(- {893\over 32}
              + {31\over 4}\,\zeta_3 \right) \logqmums
          + {71\over 12}\,\logqmums^2
          - {11\over 24}\,\logqmums^3
\nonumber\\&&\mbox{}
       + {\bar{m}^2\over q^2}\, \bigg[
          - {12617\over 48}
          + 90\,\zeta_3
          + 9\,\zeta_4
          + 15\,\zeta_5
\nonumber\\&&\mbox{\hspace{0.4cm}}
          + \left({4601\over 24}
              - 51\,\zeta_3 \right) \logqmums
          - {207\over 4}\,\logqmums^2
          + {11\over 2}\,\logqmums^3
          \bigg]
\nonumber\\&&\mbox{}
       + \left({\bar{m}^2\over q^2}\right)^{2} \, \bigg[
          - {50347\over 288}
          + {467\over 12}\,\zeta_3
          - 36\,\zeta_4
          - {95\over 2}\,\zeta_5
          + 6\,B_4
\nonumber\\&&\mbox{\hspace{0.4cm}}
          + \left(- {9503\over 48}
              + {27\over 2}\,\zeta_3 \right) \logqmms
          - 38\,\logqmms^2
          - {11\over 2}\,\logqmms^3
\nonumber\\&&\mbox{\hspace{0.4cm}}
          + \left({3005\over 24}
              + 44\,\zeta_3 
              + 136\,\logqmms
              + {33\over 2}\,\logqmms^2 \right) \logqmums
          + \left(- {231\over 8}
              - {99\over 4}\,\logqmms \right) \logqmums^2
          \bigg]
\nonumber\\&&\mbox{}
       + \left({\bar{m}^2\over q^2}\right)^{3} \, \bigg[
          - {61601\over 972}
          + {7\over 2}\,\zeta_3
          - 24\,\zeta_4
          + 20\,\zeta_5
          + 4\,B_4
\nonumber\\&&\mbox{\hspace{0.4cm}}
          + \left(- {208093\over 648}
              + 8\,\zeta_3 \right) \logqmms
          - {563\over 6}\,\logqmms^2
          - {521\over 54}\,\logqmms^3
\nonumber\\&&\mbox{\hspace{0.4cm}}
          + \left({2161\over 108}
              + {4375\over 27}\,\logqmms
              + {539\over 18}\,\logqmms^2 \right) \logqmums
          + \left({55\over 6}
              - 22\,\logqmms \right) \logqmums^2
          \bigg]
\nonumber\\&&\mbox{}
       + \left({\bar{m}^2\over q^2}\right)^{4} \, \bigg[
          {396607\over 1296}
          - {1393\over 24}\,\zeta_3
          - 36\,\zeta_4
          + 10\,\zeta_5
          + 6\,B_4
\nonumber\\&&\mbox{\hspace{0.4cm}}
          + \left(- {368443\over 864}
              + 63\,\zeta_3 \right) \logqmms
          - {57929\over 288}\,\logqmms^2
          - {803\over 24}\,\logqmms^3
\nonumber\\&&\mbox{\hspace{0.4cm}}
          + \left(- {255017\over 1728}
              + {18421\over 72}\,\logqmms
              + {495\over 8}\,\logqmms^2 \right) \logqmums
          + \left({517\over 16}
              - {165\over 4}\,\logqmms \right) \logqmums^2
          \bigg]
\bigg\}
 + \ldots\,\,, \nonumber\\
\label{eqpi2nas}
         \\[.4cm]
   \bar{\Pi}^{(2),s}_{\it l} &=& {3\over 16 \pi^2} \bigg\{
       - {511\over 36}
          + 4\,\zeta_3
          + \left({65\over 8}
              - 2\,\zeta_3 \right) \logqmums
          - {11\over 6}\,\logqmums^2
          + {1\over 6}\,\logqmums^3
\nonumber\\&&\mbox{}
       + {\bar{m}^2\over q^2}\, \bigg[
          {817\over 12}
          - 12\,\zeta_3
          + \left(- {313\over 6}
              + 12\,\zeta_3 \right) \logqmums
          + 15\,\logqmums^2
          - 2\,\logqmums^3
          \bigg]
\nonumber\\&&\mbox{}
       + \left({\bar{m}^2\over q^2}\right)^{2} \, \bigg[
          {3311\over 72}
          - {4\over 3}\,\zeta_3
          + {595\over 12}\,\logqmms
          + 10\,\logqmms^2
          + 2\,\logqmms^3
\nonumber\\&&\mbox{\hspace{0.4cm}}
          + \left(- {193\over 6}
              - 16\,\zeta_3
              - 38\,\logqmms
              - 6\,\logqmms^2 \right) \logqmums
          + \left({21\over 2}
              + 9\,\logqmms \right) \logqmums^2
          \bigg]
\nonumber\\&&\mbox{}
       + \left({\bar{m}^2\over q^2}\right)^{3} \, \bigg[
          - {8347\over 486}
          + {80\over 3}\,\zeta_3
          + {4534\over 81}\,\logqmms
          + {149\over 9}\,\logqmms^2
          + {62\over 27}\,\logqmms^3
\nonumber\\&&\mbox{\hspace{0.4cm}}
          + \left(- {311\over 27}
              - {1316\over 27}\,\logqmms
              - {98\over 9}\,\logqmms^2 \right) \logqmums
          + \left(- {10\over 3}
              + 8\,\logqmms \right) \logqmums^2
          \bigg]
\nonumber\\&&\mbox{}
       + \left({\bar{m}^2\over q^2}\right)^{4} \, \bigg[
          - {54461\over 1296}
          + 20\,\zeta_3
          + {2821\over 27}\,\logqmms
          + {287\over 8}\,\logqmms^2
          + {25\over 6}\,\logqmms^3
\nonumber\\&&\mbox{\hspace{0.4cm}}
          + \left({16723\over 432}
              - {1331\over 18}\,\logqmms
              - {45\over 2}\,\logqmms^2 \right) \logqmums
          + \left(- {47\over 4}
              + 15\,\logqmms \right) \logqmums^2
          \bigg]
\bigg\}
 + \ldots\,\,,
\label{eqpi2ls}
         \\[.4cm]
   \bar{\Pi}^{(2),s}_{\it F} &=& {3\over 16 \pi^2} \bigg\{
       - {511\over 36}
          + 4\,\zeta_3
          + \left({65\over 8}
              - 2\,\zeta_3 \right) \logqmums
          - {11\over 6}\,\logqmums^2
          + {1\over 6}\,\logqmums^3
\nonumber\\&&\mbox{}
       + {\bar{m}^2\over q^2}\, \bigg[
          {881\over 12}
          - 12\,\zeta_3
          + \left(- {385\over 6}
              + 12\,\zeta_3 \right) \logqmums
          + 15\,\logqmums^2
          - 2\,\logqmums^3
          \bigg]
\nonumber\\&&\mbox{}
       + \left({\bar{m}^2\over q^2}\right)^{2} \, \bigg[
          {9863\over 72}
          - {94\over 3}\,\zeta_3
          + {1135\over 12}\,\logqmms
          + {23\over 2}\,\logqmms^2
          + 2\,\logqmms^3
\nonumber\\&&\mbox{\hspace{0.4cm}}
          + \left(- {193\over 6}
              - 16\,\zeta_3
              - 38\,\logqmms
              - 6\,\logqmms^2 \right) \logqmums
          + \left({21\over 2}
              + 9\,\logqmms \right) \logqmums^2
          \bigg]
\nonumber\\&&\mbox{}
       + \left({\bar{m}^2\over q^2}\right)^{3} \, \bigg[
          {10757\over 486}
          - {628\over 9}\,\zeta_3
          + {4316\over 27}\,\logqmms
          + {1195\over 27}\,\logqmms^2
          + {124\over 27}\,\logqmms^3
\nonumber\\&&\mbox{\hspace{0.4cm}}
          + \left(- {311\over 27}
              - {1316\over 27}\,\logqmms
              - {98\over 9}\,\logqmms^2 \right) \logqmums
          + \left(- {10\over 3}
              + 8\,\logqmms \right) \logqmums^2
          \bigg]
\nonumber\\&&\mbox{}
       + \left({\bar{m}^2\over q^2}\right)^{4} \, \bigg[
          - {193831\over 864}
          - 187\,\zeta_3
          + {13051\over 72}\,\logqmms
          + {7699\over 72}\,\logqmms^2
          + {47\over 3}\,\logqmms^3
\nonumber\\&&\mbox{\hspace{0.4cm}}
          + \left({16723\over 432}
              - {1331\over 18}\,\logqmms
              - {45\over 2}\,\logqmms^2 \right) \logqmums
          + \left(- {47\over 4}
              + 15\,\logqmms \right) \logqmums^2
          \bigg]
\bigg\}
 + \ldots\,\,,
\label{eqpi2fs}
         \\[.4cm]
   \bar{\Pi}^{(2),s}_{\it S} &=& {3\over 16 \pi^2} \bigg\{
       {\bar{m}^2\over q^2}\, \bigg[
          {236\over 3}
          - 40\,\zeta_3
          - 20\,\zeta_5
          - 24\,\logqmums
          \bigg]
\nonumber\\&&\mbox{}
       + \left({\bar{m}^2\over q^2}\right)^{2} \, \bigg[
          - 84
          + 8\,\zeta_3
          + 160\,\zeta_5
          + \left(- 36
          + 72\,\zeta_3 \right) \logqmms
          \bigg]
\nonumber\\&&\mbox{}
       + \left({\bar{m}^2\over q^2}\right)^{3} \, \bigg[
          {37\over 8}
          - 62\,\zeta_3
          + 320\,\zeta_5
          + \left(- {3\over 4}
          - 36\,\zeta_3 \right) \logqmms
          + 33\,\logqmms^2
          + 12\,\logqmms^3
          \bigg]
\nonumber\\&&\mbox{}
       + \left({\bar{m}^2\over q^2}\right)^{4} \, \bigg[
          {178423\over 243}
          - {4472\over 9}\,\zeta_3
          + \left({22289\over 81}
          + 16\,\zeta_3 \right) \logqmms
\nonumber\\&&\mbox{\hspace{0.4cm}}
          - 26\,\logqmms^2
          - {28\over 3}\,\logqmms^3
\label{eqpi2ss}
          \bigg]
\bigg\}
 + \ldots\,\,,
\end{eqnarray}
where $\zeta$ is Riemann's zeta function with values $\zeta_2=\pi^2/6$,
$\zeta_3\approx1.20206$ $\zeta_4=\pi^4/90$ and $\zeta_5\approx
1.03693$. The constant $B_4$ is typical for three-loop tadpole
integrals and is given by $B_4\approx-1.76280$ \cite{Bro92}.  The
corresponding results in the on-shell scheme are obtained by
substituting in the combination $\bar{m}^2\bar{\Pi}(q^2)$ the
$\overline{\mbox{MS}}$ mass, $\bar{m}$, w.r.t.~the on-shell mass
\cite{GraBroGraSch90}.

For the pseudo-scalar current correlator a minor complication
arises in connection with $\gamma_5$.
While the non-singlet contribution allows the use of the anticommuting
definition, for the singlet diagram we take the prescription for
$\gamma_5$ introduced in
\cite{tHoVel72BreMai77}.
Thereby we follow the strategy outlined in
\cite{Lar93,CheKwi96}
and replace $\gamma_5$ in the vertices according to
\begin{eqnarray}
\gamma_5 &\to& \frac{i}{4!}\varepsilon_{\mu\nu\rho\sigma}
                         \gamma^{[\mu\nu\rho\sigma]},
\end{eqnarray}
where $\gamma^{[\mu\nu\rho\sigma]}$ represents the antisymmetrized
product of four $\gamma$-matrices and may be written as
\begin{eqnarray}
\gamma^{[\mu\nu\rho\sigma]} &=&
\frac{1}{4}\left(
         \gamma^{\mu}\gamma^{\nu}\gamma^{\rho}\gamma^{\sigma}
        +\gamma^{\sigma}\gamma^{\rho}\gamma^{\nu}\gamma^{\mu}
        -\gamma^{\nu}\gamma^{\rho}\gamma^{\sigma}\gamma^{\mu}
        -\gamma^{\mu}\gamma^{\sigma}\gamma^{\rho}\gamma^{\nu}
           \right).
\end{eqnarray}
Because the double-triangle diagram contains no sub-divergencies
and consequently has a finite imaginary part it is allowed to
contract the new polarization tensor with eight indices with 
the product of four metric tensors
and compute the scalar integrals in the same way as for the non-singlet
contributions.
Even more, in contrast to the scalar case $\Pi_S^{(2),p}$ is finite from the 
very beginning.

The results for the pseudo-scalar case read:
\begin{eqnarray}
   \bar{\Pi}^{(0),p} &=& {3\over 16 \pi^2} \bigg\{
       4
          - 2\,\logqmums
       + 4  {\bar{m}^2\over q^2}\logqmums
       + \left({\bar{m}^2\over q^2}\right)^{2} \, \bigg[
          - 2
          + 4\,\logqmms
          \bigg]
\nonumber\\&&\mbox{}
       + \left({\bar{m}^2\over q^2}\right)^{3} \, \bigg[
          - {20\over 3}
          + 8\,\logqmms
          \bigg]
       + \left({\bar{m}^2\over q^2}\right)^{4} \, \bigg[
          - {59\over 3}
          + 20\,\logqmms
          \bigg]
\bigg\}
 + \ldots\,\,,
\label{eqpi0p}
         \\[.4cm]
   \bar{\Pi}^{(1),p} &=& {3\over 16 \pi^2} \bigg\{
       {131\over 8}
          - 6\,\zeta_3
          - {17\over 2}\,\logqmums
          + {3\over 2}\,\logqmums^2
       + {\bar{m}^2\over q^2}\, \bigg[
          - 6
          + 12\,\zeta_3
          + 4\,\logqmums
          - 6\,\logqmums^2
          \bigg]
\nonumber\\&&\mbox{}
       + \left({\bar{m}^2\over q^2}\right)^{2} \, \bigg[
          4
          + 27\,\logqmms
          + 6\,\logqmms^2
          + \left( 15
          - 18\,\logqmms\right) \logqmums
          \bigg]
\nonumber\\&&\mbox{}
       + \left({\bar{m}^2\over q^2}\right)^{3} \, \bigg[
          - {515\over 9}
          + {628\over 9}\,\logqmms
          + {82\over 3}\,\logqmms^2
          + \left( 52
          - 48\,\logqmms\right) \logqmums
          \bigg]
\nonumber\\&&\mbox{}
       + \left({\bar{m}^2\over q^2}\right)^{4} \, \bigg[
          - {36985\over 144}
          + {3101\over 18}\,\logqmms
          + {539\over 6}\,\logqmms^2
          + \left( {355\over 2}
          - 150\,\logqmms\right) \logqmums
          \bigg]
\bigg\}
 + \ldots\,\,,
\nonumber\\
\label{eqpi1p}
         \\[.4cm]
   \bar{\Pi}^{(2),p}_{\it A} &=& {3\over 16 \pi^2} \bigg\{
       {1613\over 64}
          - 24\,\zeta_3
          + {9\over 4}\,\zeta_4
          + 15\,\zeta_5
          + \left( - {691\over 32}
          + {9\over 2}\,\zeta_3\right) \logqmums
          + {105\over 16}\,\logqmums^2
          - {3\over 4}\,\logqmums^3
\nonumber\\&&\mbox{}
       + {\bar{m}^2\over q^2}\, \bigg[
          {1697\over 48}
          + 54\,\zeta_3
          - 6\,\zeta_4
          - 70\,\zeta_5
          + \left( {13\over 8}
          - 24\,\zeta_3\right) \logqmums
          - {27\over 4}\,\logqmums^2
          + 6\,\logqmums^3
          \bigg]
\nonumber\\&&\mbox{}
       + \left({\bar{m}^2\over q^2}\right)^{2} \, \bigg[
          - {885\over 32}
          + 104\,\zeta_3
          - 24\,\zeta_4
          + 15\,\zeta_5
          + 4\,B_4
          + \left( {1253\over 16}
          - 33\,\zeta_3\right) \logqmms
\nonumber\\&&\mbox{\hspace{0.4cm}}
          + {129\over 4}\,\logqmms^2
          + 6\,\logqmms^3
          + \left( {195\over 8}
          - {423\over 4}\,\logqmms
          - 27\,\logqmms^2\right) \logqmums
          + \left( - {189\over 4}
          + {81\over 2}\,\logqmms\right) \logqmums^2
          \bigg]
\nonumber\\&&\mbox{}
       + \left({\bar{m}^2\over q^2}\right)^{3} \, \bigg[
          - {421657\over 1944}
          + 135\,\zeta_3
          - 48\,\zeta_4
          + 80\,\zeta_5
          + 8\,B_4
          + \left( {4481\over 27}
          + 52\,\zeta_3\right) \logqmms
\nonumber\\&&\mbox{\hspace{0.4cm}}
          + {839\over 6}\,\logqmms^2
          + {1223\over 27}\,\logqmms^3
          + \left( {909\over 2}
          - {1028\over 3}\,\logqmms
          - 164\,\logqmms^2\right) \logqmums
\nonumber\\&&\mbox{\hspace{0.4cm}}
          + \left( - 192
          + 144\,\logqmms\right) \logqmums^2
          \bigg]
\nonumber\\&&\mbox{}
       + \left({\bar{m}^2\over q^2}\right)^{4} \, \bigg[
          - {159609\over 128}
          - {50\over 3}\,\zeta_3
          - 120\,\zeta_4
          + 280\,\zeta_5
          + 20\,B_4
\nonumber\\&&\mbox{\hspace{0.4cm}}
          + \left( {751555\over 1296}
          + 203\,\zeta_3\right) \logqmms
          + {534067\over 864}\,\logqmms^2
          + {38753\over 216}\,\logqmms^3
\nonumber\\&&\mbox{\hspace{0.4cm}}
          + \left( {70621\over 32}
          - {3124\over 3}\,\logqmms
          - {2695\over 4}\,\logqmms^2\right) \logqmums
          + \left( - {6225\over 8}
          + {1125\over 2}\,\logqmms\right) \logqmums^2
          \bigg]
\bigg\}
 + \ldots\,\,, \nonumber \\
\label{eqpi2ap}
         \\[.4cm]
   \bar{\Pi}^{(2),p}_{\it NA} &=& {3\over 16 \pi^2} \bigg\{
       {14419\over 288}
          - {75\over 4}\,\zeta_3
          - {9\over 8}\,\zeta_4
          - {5\over 2}\,\zeta_5
          + \left( - {893\over 32}
          + {31\over 4}\,\zeta_3\right) \logqmums
          + {71\over 12}\,\logqmums^2
          - {11\over 24}\,\logqmums^3
\nonumber\\&&\mbox{}
       + {\bar{m}^2\over q^2}\, \bigg[
          - {25\over 144}
          + {44\over 3}\,\zeta_3
          + 3\,\zeta_4
          + 5\,\zeta_5
          + \left( {153\over 8}
          - 17\,\zeta_3\right) \logqmums
\nonumber\\&&\mbox{\hspace{0.4cm}}
          - {119\over 12}\,\logqmums^2
          + {11\over 6}\,\logqmums^3
          \bigg]
\nonumber\\&&\mbox{}
       + \left({\bar{m}^2\over q^2}\right)^{2} \, \bigg[
          - {1417\over 96}
          + {49\over 4}\,\zeta_3
          + 12\,\zeta_4
          + {5\over 2}\,\zeta_5
          - 2\,B_4
          + \left( {3925\over 48}
          + {15\over 2}\,\zeta_3\right) \logqmms
\nonumber\\&&\mbox{\hspace{0.4cm}}
          + {38\over 3}\,\logqmms^2
          + {11\over 6}\,\logqmms^3
          + \left( {397\over 24}
          - 49\,\logqmms
          - {11\over 2}\,\logqmms^2\right) \logqmums
          + \left( - {55\over 8}
          + {33\over 4}\,\logqmms\right) \logqmums^2
          \bigg]
\nonumber\\&&\mbox{}
       + \left({\bar{m}^2\over q^2}\right)^{3} \, \bigg[
          - {134501\over 972}
          - {63\over 2}\,\zeta_3
          + 24\,\zeta_4
          - 20\,\zeta_5
          - 4\,B_4
          + \left( {170609\over 648}
          - 40\,\zeta_3\right) \logqmms
\nonumber\\&&\mbox{\hspace{0.4cm}}
          + {497\over 6}\,\logqmms^2
          + {505\over 54}\,\logqmms^3
          + \left( {13231\over 108}
          - {3473\over 27}\,\logqmms
          - {451\over 18}\,\logqmms^2\right) \logqmums
\nonumber\\&&\mbox{\hspace{0.4cm}}
          + \left( - {143\over 6}
          + 22\,\logqmms\right) \logqmums^2
          \bigg]
\nonumber\\&&\mbox{}
       + \left({\bar{m}^2\over q^2}\right)^{4} \, \bigg[
          - {3008885\over 3888}
          + {823\over 24}\,\zeta_3
          + 60\,\zeta_4
          - 70\,\zeta_5
          - 10\,B_4
\nonumber\\&&\mbox{\hspace{0.4cm}}
          + \left( {1489679\over 2592}
          - 145\,\zeta_3\right) \logqmms
          + {74807\over 288}\,\logqmms^2
          + {8941\over 216}\,\logqmms^3
\nonumber\\&&\mbox{\hspace{0.4cm}}
          + \left( {820055\over 1728}
          - {77761\over 216}\,\logqmms
          - {5929\over 72}\,\logqmms^2\right) \logqmums
          + \left( - {3905\over 48}
          + {275\over 4}\,\logqmms\right) \logqmums^2
          \bigg]
\bigg\}
 + \ldots\,\,, \nonumber\\
\label{eqpi2nap}
         \\[.4cm]
   \bar{\Pi}^{(2),p}_{\it l} &=& {3\over 16 \pi^2} \bigg\{
       - {511\over 36}
          + 4\,\zeta_3
          + \left( {65\over 8}
          - 2\,\zeta_3\right) \logqmums
          - {11\over 6}\,\logqmums^2
          + {1\over 6}\,\logqmums^3
\nonumber\\&&\mbox{}
       + {\bar{m}^2\over q^2}\, \bigg[
          - {31\over 36}
          - {4\over 3}\,\zeta_3
          + \left( - {9\over 2}
          + 4\,\zeta_3\right) \logqmums
          + {7\over 3}\,\logqmums^2
          - {2\over 3}\,\logqmums^3
          \bigg]
\nonumber\\&&\mbox{}
       + \left({\bar{m}^2\over q^2}\right)^{2} \, \bigg[
          {53\over 24}
          - 12\,\zeta_3
          - {257\over 12}\,\logqmms
          - {10\over 3}\,\logqmms^2
          - {2\over 3}\,\logqmms^3
\nonumber\\&&\mbox{\hspace{0.4cm}}
          + \left( - {17\over 6}
          + 14\,\logqmms
          + 2\,\logqmms^2\right) \logqmums
          + \left( {5\over 2}
          - 3\,\logqmms\right) \logqmums^2
          \bigg]
\nonumber\\&&\mbox{}
       + \left({\bar{m}^2\over q^2}\right)^{3} \, \bigg[
          {12551\over 486}
          - {16\over 3}\,\zeta_3
          - {4970\over 81}\,\logqmms
          - {133\over 9}\,\logqmms^2
          - {46\over 27}\,\logqmms^3
\nonumber\\&&\mbox{\hspace{0.4cm}}
          + \left( - {905\over 27}
          + {988\over 27}\,\logqmms
          + {82\over 9}\,\logqmms^2\right) \logqmums
          + \left( {26\over 3}
          - 8\,\logqmms\right) \logqmums^2
          \bigg]
\nonumber\\&&\mbox{}
       + \left({\bar{m}^2\over q^2}\right)^{4} \, \bigg[
          {573913\over 3888}
          - {4\over 3}\,\zeta_3
          - {13547\over 81}\,\logqmms
          - {1315\over 24}\,\logqmms^2
          - {287\over 54}\,\logqmms^3
\nonumber\\&&\mbox{\hspace{0.4cm}}
          + \left( - {58285\over 432}
          + {5351\over 54}\,\logqmms
          + {539\over 18}\,\logqmms^2\right) \logqmums
          + \left( {355\over 12}
          - 25\,\logqmms\right) \logqmums^2
          \bigg]
\bigg\}
 + \ldots\,\,,
\label{eqpi2lp}
         \\[.4cm]
   \bar{\Pi}^{(2),p}_{\it F} &=& {3\over 16 \pi^2} \bigg\{
       - {511\over 36}
          + 4\,\zeta_3
          + \left( {65\over 8}
          - 2\,\zeta_3\right) \logqmums
          - {11\over 6}\,\logqmums^2
          + {1\over 6}\,\logqmums^3
\nonumber\\&&\mbox{}
       + {\bar{m}^2\over q^2}\, \bigg[
          {161\over 36}
          - {4\over 3}\,\zeta_3
          + \left( - {33\over 2}
          + 4\,\zeta_3\right) \logqmums
          + {7\over 3}\,\logqmums^2
          - {2\over 3}\,\logqmums^3
          \bigg]
\nonumber\\&&\mbox{}
       + \left({\bar{m}^2\over q^2}\right)^{2} \, \bigg[
          {701\over 24}
          + 6\,\zeta_3
          - {293\over 12}\,\logqmms
          - {11\over 6}\,\logqmms^2
          - {2\over 3}\,\logqmms^3
\nonumber\\&&\mbox{\hspace{0.4cm}}
          + \left( - {17\over 6}
          + 14\,\logqmms
          + 2\,\logqmms^2\right) \logqmums
          + \left( {5\over 2}
          - 3\,\logqmms\right) \logqmums^2
          \bigg]
\nonumber\\&&\mbox{}
       + \left({\bar{m}^2\over q^2}\right)^{3} \, \bigg[
          {43319\over 486}
          + {596\over 9}\,\zeta_3
          - {3064\over 27}\,\logqmms
          - {623\over 27}\,\logqmms^2
          - {92\over 27}\,\logqmms^3
\nonumber\\&&\mbox{\hspace{0.4cm}}
          + \left( - {905\over 27}
          + {988\over 27}\,\logqmms
          + {82\over 9}\,\logqmms^2\right) \logqmums
          + \left( {26\over 3}
          - 8\,\logqmms\right) \logqmums^2
          \bigg]
\nonumber\\&&\mbox{}
       + \left({\bar{m}^2\over q^2}\right)^{4} \, \bigg[
          {3892289\over 7776}
          + {2333\over 9}\,\zeta_3
          - {20533\over 72}\,\logqmms
          - {24871\over 216}\,\logqmms^2
          - {545\over 27}\,\logqmms^3
\nonumber\\&&\mbox{\hspace{0.4cm}}
          + \left( - {58285\over 432}
          + {5351\over 54}\,\logqmms
          + {539\over 18}\,\logqmms^2\right) \logqmums
          + \left( {355\over 12}
          - 25\,\logqmms\right) \logqmums^2
          \bigg]
\bigg\}
 + \ldots\,\,,
\label{eqpi2fp}
         \\[.4cm]
   \bar{\Pi}^{(2),p}_{\it S} &=& {3\over 16 \pi^2} \bigg\{
       {\bar{m}^2\over q^2}\, \bigg[
          - 16\,\zeta_3
          - 20\,\zeta_5
          \bigg]
\nonumber\\&&\mbox{}
       + \left({\bar{m}^2\over q^2}\right)^{2} \, \bigg[
          - 44
          + 24\,\zeta_3
          + \left( - 12
          - 72\,\zeta_3\right) \logqmms
          \bigg]
\nonumber\\&&\mbox{}
       + \left({\bar{m}^2\over q^2}\right)^{3} \, \bigg[
          {221\over 8}
          + 114\,\zeta_3
          + \left( - {363\over 4}
          - 36\,\zeta_3\right) \logqmms
          - 63\,\logqmms^2
          - 12\,\logqmms^3
          \bigg]
\nonumber\\&&\mbox{}
       + \left({\bar{m}^2\over q^2}\right)^{4} \, \bigg[
          {68146\over 243}
          + {1288\over 9}\,\zeta_3
          + \left( {7727\over 81}
          - 80\,\zeta_3\right) \logqmms
          - 86\,\logqmms^2
          - 44\,\logqmms^3
          \bigg]
\bigg\}
 + \ldots\,\,, \nonumber\\
\label{eqpi2sp}
\end{eqnarray}
The results for the one- and two-loop 
case have been checked against the analytical exact expressions
for the polarization functions
\cite{DjoGam95}
in the limit of large external momentum $q$.


\section{\label{sechtt}The decays \boldmath{$H\to t\bar{t}$}
 and \boldmath{$A\to t\bar{t}$}}

As an application of the results of the previous section we 
consider the decay of a scalar and pseudo-scalar Higgs boson 
into a top-quark pair. 
Thereby we neglect all light quark masses and denote the 
$\overline{\mbox{MS}}$ top quark mass by $m_t$ and the on-shell one
by $M_t$.
Let us in a first step review the Born result and the first order
QCD corrections. Taking the imaginary part of 
Eqs.~(\ref{eqpi0s}) and (\ref{eqpi1s})
and transforming the combination $m_t^2\bar{R}(s)$ 
into the on-shell scheme
\cite{GraBroGraSch90}
one arrives at ($\logmsos \equiv \ln(M_t^2/s)$)
\begin{eqnarray}
   R^{(0),s} &=& 3\, \bigg\{
       1
       - 6 {M_t^2\over s}
       + 6 \left({M_t^2\over s}\right)^{2}
       + 4 \left({M_t^2\over s}\right)^{3}
       + 6 \left({M_t^2\over s}\right)^{4}
\bigg\}
 + \ldots\,\,,
         \\[.4cm]
   R^{(1),s} &=&  3\, \bigg\{
       {9\over 4}
          + {3\over 2}\,\logmsos
       + {M_t^2\over s}\, \bigg[
          - 6
          - 18\,\logmsos
          \bigg]
       + \left({M_t^2\over s}\right)^{2} \, \bigg[
          - 33
          + 36\,\logmsos
          \bigg]
\nonumber\\&&\mbox{}
       + \left({M_t^2\over s}\right)^{3} \, \bigg[
          {280\over 9}
          + {46\over 3}\,\logmsos
          \bigg]
       + \left({M_t^2\over s}\right)^{4} \, \bigg[
          {146\over 3}
          + {45\over 2}\,\logmsos
          \bigg]
\bigg\}
 + \ldots\,\,,
\end{eqnarray}
for the scalar case where $\mu^2=s$ is chosen. For the 
pseudo-scalar current we get from Eqs.~(\ref{eqpi0p}) and (\ref{eqpi1p}):
\begin{eqnarray}
   R^{(0),p} &=&  3 \, \bigg\{
       1
       - 2 {M_t^2\over s}
       - 2 \left({M_t^2\over s}\right)^{2}
       - 4 \left({M_t^2\over s}\right)^{3}
       - 10 \left({M_t^2\over s}\right)^{4}
\bigg\}
 + \ldots\,\,,
         \\[.4cm]
   R^{(1),p} &=&  3 \, \bigg\{
       {9\over 4}
          + {3\over 2}\,\logmsos
       + {M_t^2\over s} \, \bigg[
          6
          - 6\,\logmsos
          \bigg]
       + \left({M_t^2\over s}\right)^{2} \, \bigg[
          - 9
          - 12\,\logmsos
          \bigg]
\nonumber\\&&\mbox{}
       + \left({M_t^2\over s}\right)^{3} \, \bigg[
          - {260\over 9}
          - {62\over 3}\,\logmsos
          \bigg]
       + \left({M_t^2\over s}\right)^{4} \, \bigg[
          - {674\over 9}
          - {361\over 6}\,\logmsos
          \bigg]
\bigg\}
 + \ldots\,.
\end{eqnarray}

\begin{figure}[htb]
 \begin{center}
 \begin{tabular}{cc}
   \leavevmode
   \epsfxsize=6.5cm
   \epsffile[110 270 480 560]{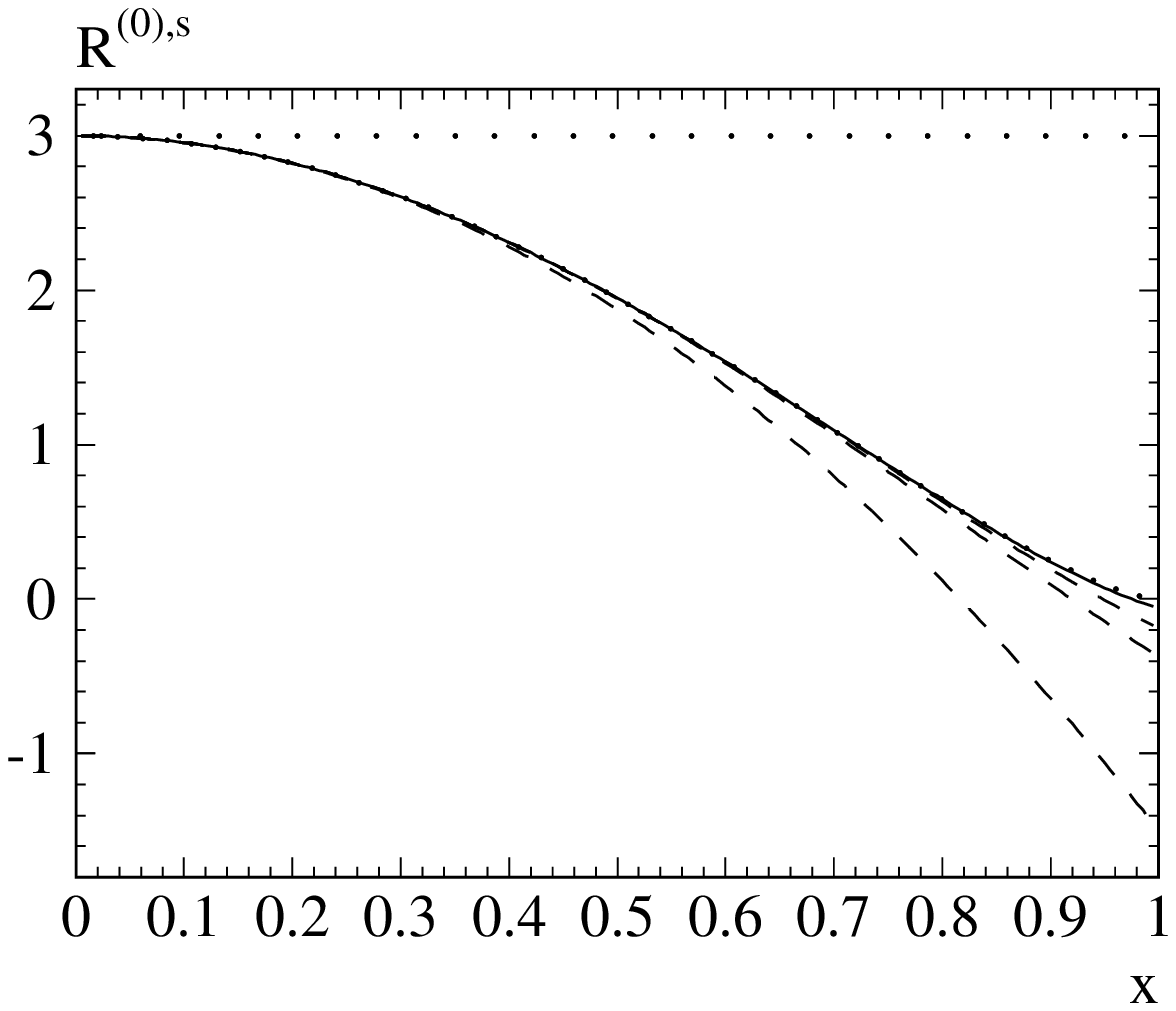 }
   &
   \epsfxsize=6.5cm
   \epsffile[110 270 480 560]{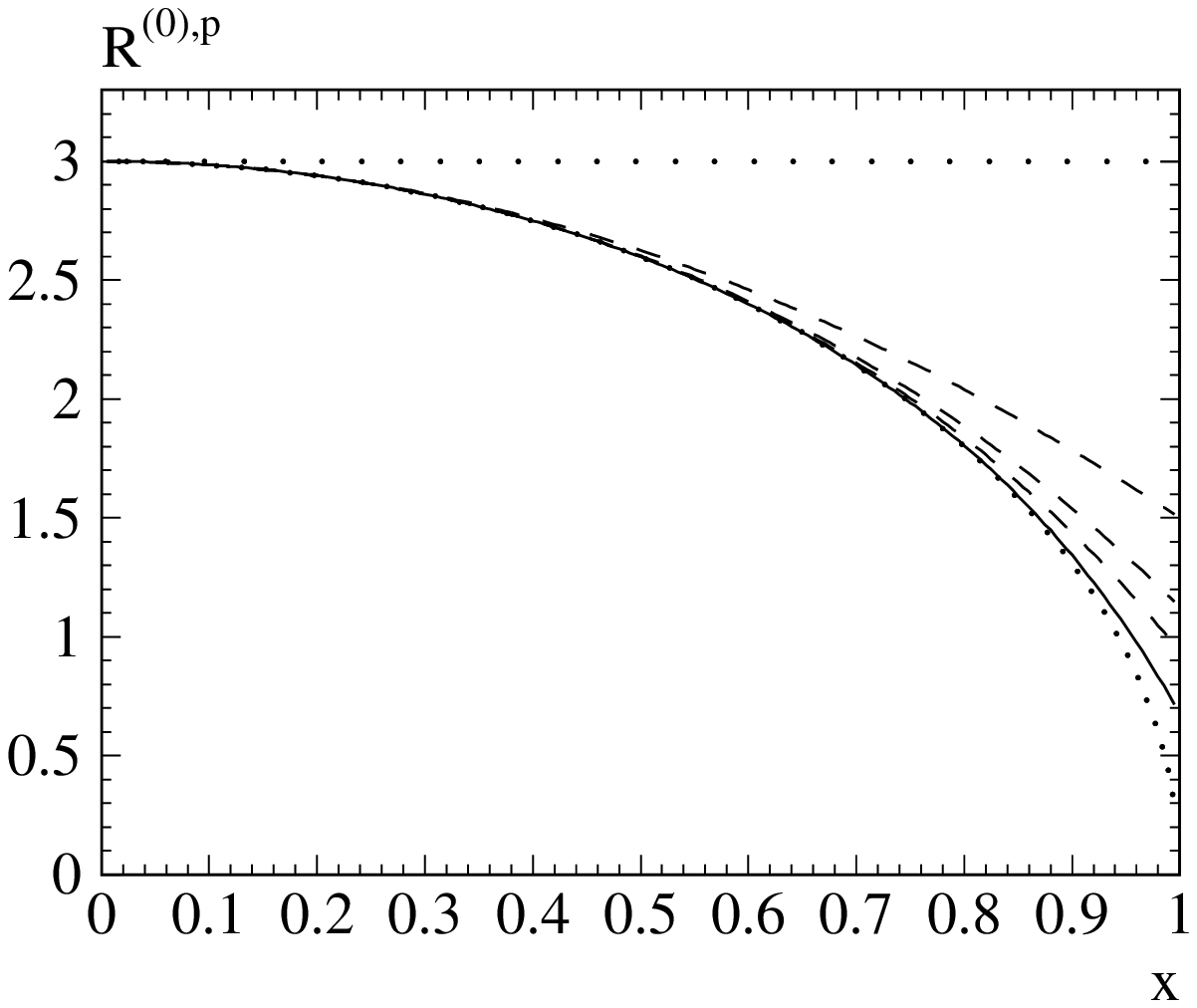 }
\\
   \epsfxsize=6.5cm
   \epsffile[110 270 480 560]{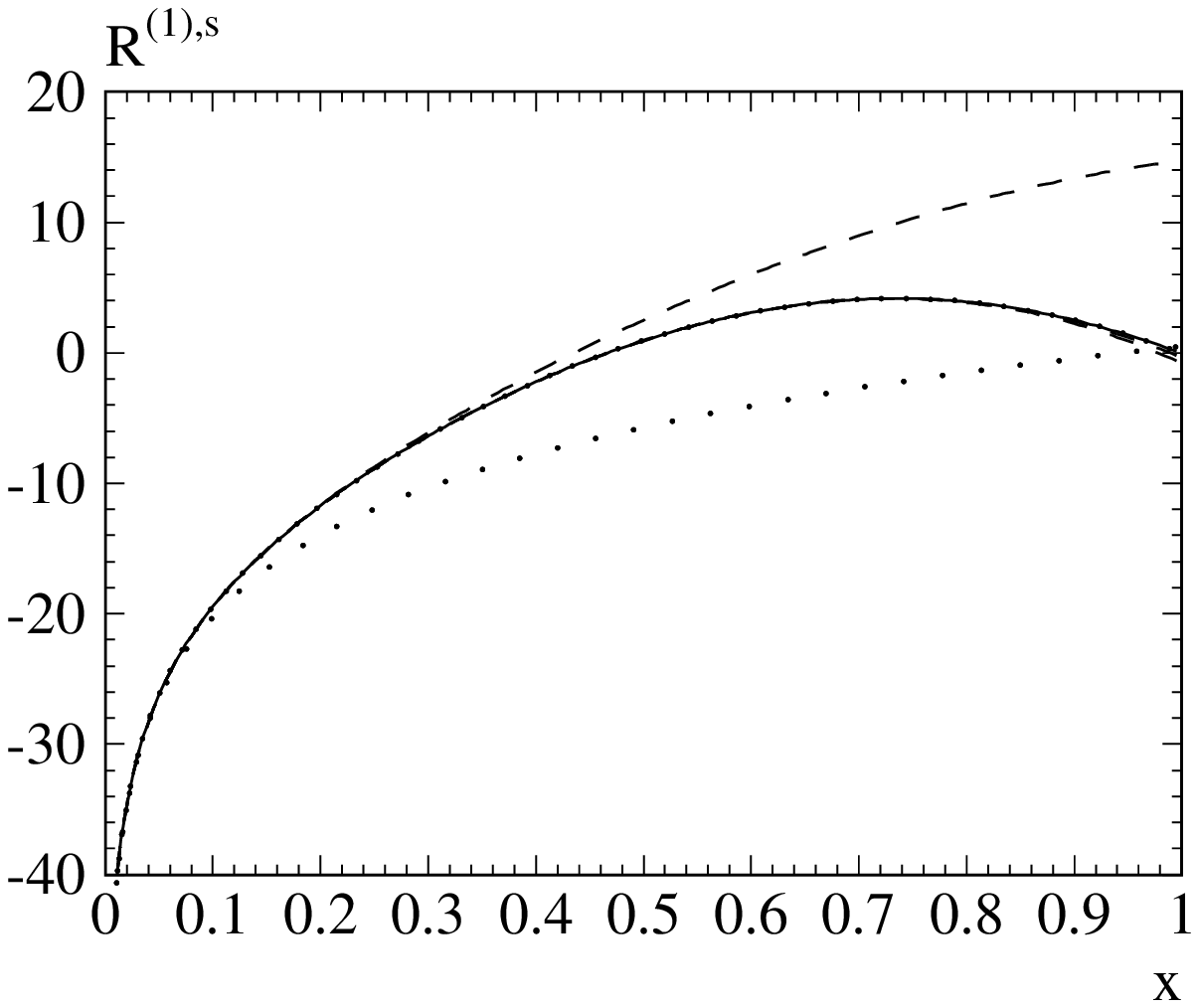 }
   &
   \epsfxsize=6.5cm
   \epsffile[110 270 480 560]{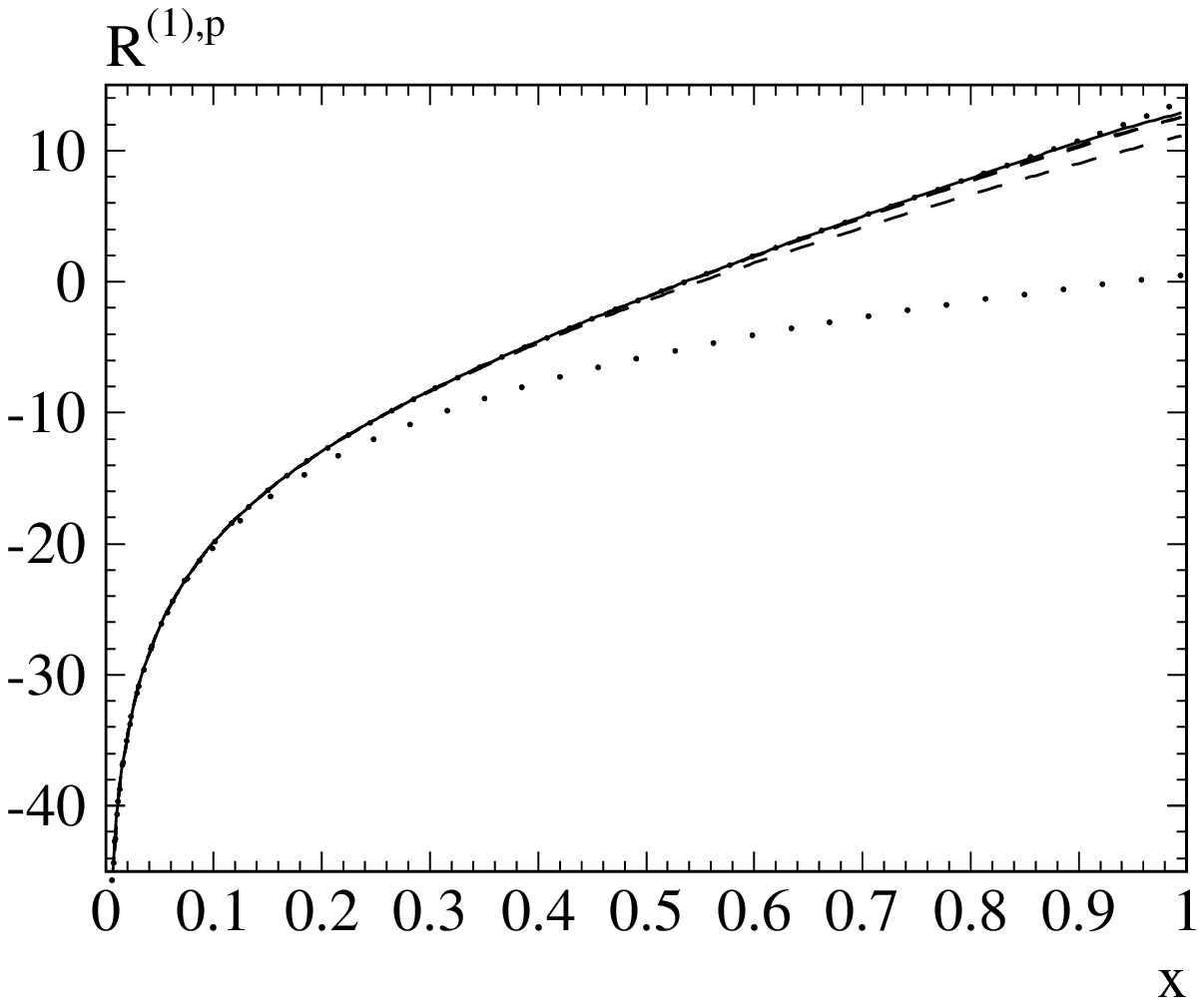 }
 \end{tabular}
 \caption{\label{figsp01} 
   $R^{(0),\delta}$ and $R^{(1),\delta}$, $\delta=s,p$, plotted against 
   $x=2M_t/\protect\sqrt{s}$.
   Successively higher order terms in $(M_t^2/s)^n$:
   Wide dots: $n=0$; dashed: $n=1,2,3$;
   solid: $n=4$;
   narrow dots: exact.
}
 \end{center}
\end{figure}

In Fig.~\ref{figsp01} the results are plotted together with the exact
expressions.  For the abscissa the variable $x=2M_t/\sqrt{s}$ is chosen.
Already the approximations including only the quadratic terms provide a
good agreement up to $x\approx0.4$. It is, however, dramatically
improved by incorporating higher terms in $M_t^2/s$, leading to an
excellent approximation almost up to $x=1$ for $R^{(0),s}, R^{(1),s}$
and $R^{(1),p}$ and up to $x\approx0.85$ for $R^{(0),p}$.  This serves
as a strong motivation to evaluate higher order mass terms for the
${\cal O}(\alpha_s^2)$ corrections.

At ${\cal O}(\alpha_s^2)$ the results for $R^{(2),s}(s)$, separated into 
the contributions from the different colour factors, read in the
on-shell scheme:
\begin{eqnarray}
   R^{(2),s}_{\it A} &=&  3\, \bigg\{
       {109\over 32}
          + \left(- 6
             + 6\,\ln 2 \right)\,\zeta_2
          - {15\over 4}\,\zeta_3
          + {57\over 16}\,\logmsos
          + {9\over 8}\,\logmsos^2
\nonumber\\&&\mbox{}
       + {M_t^2\over s}\, \bigg[
          - {147\over 4}
          + \left( 99
              - 72\,\ln 2 \right)\,\zeta_2
          + 54\,\zeta_3
          - {81\over 4}\,\logmsos
          - 27\,\logmsos^2
          \bigg]
\nonumber\\&&\mbox{}
       + \left({M_t^2\over s}\right)^{2} \, \bigg[
          {189\over 2}
          + \left(- 243
             + 108\,\ln 2 \right)\,\zeta_2
          - {297\over 2}\,\zeta_3
          - {531\over 4}\,\logmsos
          + {351\over 4}\,\logmsos^2
          \bigg]
\nonumber\\&&\mbox{}
       + \left({M_t^2\over s}\right)^{3} \, \bigg[
          - {7588\over 27}
          + \left( - {775\over 9}
              + 96\,\ln 2 \right)\,\zeta_2
          + 34\,\zeta_3
          + {579\over 2}\,\logmsos
          + {235\over 18}\,\logmsos^2
          \bigg]
\nonumber\\&&\mbox{}
       + \left({M_t^2\over s}\right)^{4} \, \bigg[
          {211615\over 5184}
          + \left( - {4385\over 72}
              + 180\,\ln 2 \right)\,\zeta_2
          + {83\over 2}\,\zeta_3
\nonumber\\&&\mbox{\hspace{.5cm}}
          + {332551\over 864}\,\logmsos
          - {3715\over 144}\,\logmsos^2
          \bigg]
\bigg\}
 + \ldots\,\,,
         \\[.4cm]
   R^{(2),s}_{\it NA} &=& 3\, \bigg\{
       {49\over 6}
          + \left( - {3\over 8}
              - 3\,\ln 2 \right)\,\zeta_2
          - {25\over 8}\,\zeta_3
          + {185\over 48}\,\logmsos
          - {11\over 16}\,\logmsos^2
\nonumber\\&&\mbox{}
       + {M_t^2\over s}\, \bigg[
          - {317\over 12}
          + \left( {9\over 2}
              + 36\,\ln 2 \right)\,\zeta_2
          + {33\over 2}\,\zeta_3
          - {185\over 4}\,\logmsos
          + {33\over 4}\,\logmsos^2
          \bigg]
\nonumber\\&&\mbox{}
       + \left({M_t^2\over s}\right)^{2} \, \bigg[
          - {3253\over 48}
          + \left( - {93\over 4}
              - 54\,\ln 2 \right)\,\zeta_2
          - {61\over 4}\,\zeta_3
          + {795\over 8}\,\logmsos
          - {99\over 8}\,\logmsos^2
          \bigg]
\nonumber\\&&\mbox{}
       + \left({M_t^2\over s}\right)^{3} \, \bigg[
          {75139\over 1296}
          + \left( {98\over 9}
              - 48\,\ln 2 \right)\,\zeta_2
          + 8\,\zeta_3
          + {1319\over 27}\,\logmsos
          - {23\over 2}\,\logmsos^2
          \bigg]
\nonumber\\&&\mbox{}
       + \left({M_t^2\over s}\right)^{4} \, \bigg[
          {391963\over 3456}
          + \left( - {17\over 2}
              - 90\,\ln 2 \right)\,\zeta_2
          - 9\,\zeta_3
\nonumber\\&&\mbox{\hspace{.5cm}}
          + {1357\over 32}\,\logmsos
          - {11\over 8}\,\logmsos^2
          \bigg]
\bigg\}
 + \ldots\,\,, 
         \\[.4cm]
   R^{(2),s}_{\it l} &=&  3\, \bigg\{
       - {31\over 12}
          + {3\over 2}\,\zeta_2
          + \zeta_3
          - {13\over 12}\,\logmsos
          + {1\over 4}\,\logmsos^2
\nonumber\\&&\mbox{}
       + {M_t^2\over s}\, \bigg[
          {25\over 3}
          - 18\,\zeta_2
          - 6\,\zeta_3
          + 13\,\logmsos
          - 3\,\logmsos^2
          \bigg]
\nonumber\\&&\mbox{}
       + \left({M_t^2\over s}\right)^{2} \, \bigg[
          {215\over 12}
          + 33\,\zeta_2
          + 8\,\zeta_3
          - {57\over 2}\,\logmsos
          + {9\over 2}\,\logmsos^2
          \bigg]
\nonumber\\&&\mbox{}
       + \left({M_t^2\over s}\right)^{3} \, \bigg[
          {233\over 162}
          + {128\over 9}\,\zeta_2
          - {679\over 27}\,\logmsos
          + 6\,\logmsos^2
          \bigg]
\nonumber\\&&\mbox{}
       + \left({M_t^2\over s}\right)^{4} \, \bigg[
          - {23519\over 864}
          + 20\,\zeta_2
          - {2419\over 72}\,\logmsos
          + {25\over 2}\,\logmsos^2
          \bigg]
\bigg\}
 + \ldots\,\,,
         \\[.4cm]
   R^{(2),s}_{\it F} &=&  3\, \bigg\{
       - {13\over 12}
          - {3\over 2}\,\zeta_2
          + \zeta_3
          - {13\over 12}\,\logmsos
          + {1\over 4}\,\logmsos^2
\nonumber\\&&\mbox{}
       + {M_t^2\over s}\, \bigg[
          - {11\over 3}
          + 18\,\zeta_2
          - 6\,\zeta_3
          + 13\,\logmsos
          - 3\,\logmsos^2
          \bigg]
\nonumber\\&&\mbox{}
       + \left({M_t^2\over s}\right)^{2} \, \bigg[
          {269\over 12}
          - 21\,\zeta_2
          + 8\,\zeta_3
          - 27\,\logmsos
          + {9\over 2}\,\logmsos^2
          \bigg]
\nonumber\\&&\mbox{}
       + \left({M_t^2\over s}\right)^{3} \, \bigg[
          - {53\over 2}
          - {242\over 9}\,\zeta_2
          + {23\over 9}\,\logmsos
          + {23\over 9}\,\logmsos^2
          \bigg]
\nonumber\\&&\mbox{}
       + \left({M_t^2\over s}\right)^{4} \, \bigg[
          - {17809\over 864}
          - {71\over 2}\,\zeta_2
          + {899\over 24}\,\logmsos
          - {19\over 4}\,\logmsos^2
          \bigg]
\bigg\}
 + \ldots\,\,,
         \\[.4cm]
   R^{(2),s}_{\it S} &=&  3\, \bigg\{
       12 {M_t^2\over s}
       + \left({M_t^2\over s}\right)^{2} \, \bigg[
          18
          - 36\,\zeta_3
          \bigg]
\nonumber\\&&\mbox{}
       + \left({M_t^2\over s}\right)^{3} \, \bigg[
          {3\over 8}
          + 36\,\zeta_2
          + 18\,\zeta_3
          + 33\,\logmsos
          - 18\,\logmsos^2
          \bigg]
\nonumber\\&&\mbox{}
       + \left({M_t^2\over s}\right)^{4} \, \bigg[
          - {22289\over 162}
          - 28\,\zeta_2
          - 8\,\zeta_3
          - 26\,\logmsos
          + 14\,\logmsos^2
          \bigg]
\bigg\}
 + \ldots\,\,.
\end{eqnarray}
In the pseudo-scalar case we get:
\begin{eqnarray}
   R^{(2),p}_{\it A} &=&  3 \, \bigg\{
       {109\over 32}
          + \left( - 6
              + 6\,\ln 2 \right)\,\zeta_2
          - {15\over 4}\,\zeta_3
          + {57\over 16}\,\logmsos
          + {9\over 8}\,\logmsos^2
\nonumber\\&&\mbox{}
       + {M_t^2\over s} \, \bigg[
          - {21\over 4}
          + \left( 33
              - 24\,\ln 2 \right)\,\zeta_2
          + 18\,\zeta_3
          + {69\over 4}\,\logmsos
          - 9\,\logmsos^2
          \bigg]
\nonumber\\&&\mbox{}
       + \left({M_t^2\over s}\right)^{2} \, \bigg[
          50
          + \left( 81
          - 36\,\ln 2 \right)\,\zeta_2
          + {51\over 2}\,\zeta_3
          - {195\over 4}\,\logmsos
          - {117\over 4}\,\logmsos^2
          \bigg]
\nonumber\\&&\mbox{}
       + \left({M_t^2\over s}\right)^{3} \, \bigg[
          {3049\over 54}
          + \left( {1223\over 9}
              - 96\,\ln 2 \right)\,\zeta_2
          - 2\,\zeta_3
          - {459\over 2}\,\logmsos
          - {683\over 18}\,\logmsos^2
          \bigg]
\nonumber\\&&\mbox{}
       + \left({M_t^2\over s}\right)^{4} \, \bigg[
          - {672113\over 5184}
          + \left( {28223\over 72}
              - 300\,\ln 2 \right)\,\zeta_2
          - {53\over 2}\,\zeta_3
\nonumber\\&&\mbox{\hspace{.5cm}}
          - {518105\over 864}\,\logmsos
          - {14723\over 144}\,\logmsos^2
          \bigg]
\bigg\}
 + \ldots\,\,,
         \\[.4cm]
   R^{(2),p}_{\it NA} &=&  3 \, \bigg\{
       {49\over 6}
          + \left( - {3\over 8}
              - 3\,\ln 2 \right)\,\zeta_2
          - {25\over 8}\,\zeta_3
          + {185\over 48}\,\logmsos
          - {11\over 16}\,\logmsos^2
\nonumber\\&&\mbox{}
       + {M_t^2\over s} \, \bigg[
          {163\over 12}
          + \left( {3\over 2}
              + 12\,\ln 2 \right)\,\zeta_2
          + {11\over 2}\,\zeta_3
          - {185\over 12}\,\logmsos
          + {11\over 4}\,\logmsos^2
          \bigg]
\nonumber\\&&\mbox{}
       + \left({M_t^2\over s}\right)^{2} \, \bigg[
          - {231\over 16}
          + \left( {31\over 4}
              + 18\,\ln 2 \right)\,\zeta_2
          - {33\over 4}\,\zeta_3
          - {839\over 24}\,\logmsos
          + {33\over 8}\,\logmsos^2
          \bigg]
\nonumber\\&&\mbox{}
       + \left({M_t^2\over s}\right)^{3} \, \bigg[
          - {130007\over 1296}
          + \left( {26\over 9}
              + 48\,\ln 2 \right)\,\zeta_2
          + 8\,\zeta_3
          - {1165\over 27}\,\logmsos
          + {19\over 2}\,\logmsos^2
          \bigg]
\nonumber\\&&\mbox{}
       + \left({M_t^2\over s}\right)^{4} \, \bigg[
          - {2439823\over 10368}
          + \left( {601\over 18}
              + 150\,\ln 2 \right)\,\zeta_2
          + 35\,\zeta_3
\nonumber\\&&\mbox{\hspace{.5cm}}
          - {97601\over 864}\,\logmsos
          + {323\over 24}\,\logmsos^2
          \bigg]
\bigg\}
 + \ldots\,\,,
         \\[.4cm]
   R^{(2),p}_{\it l} &=&  3 \, \bigg\{
       - {31\over 12}
          + {3\over 2}\,\zeta_2
          + \zeta_3
          - {13\over 12}\,\logmsos
          + {1\over 4}\,\logmsos^2
\nonumber\\&&\mbox{}
       + {M_t^2\over s} \, \bigg[
          - {11\over 3}
          - 6\,\zeta_2
          - 2\,\zeta_3
          + {13\over 3}\,\logmsos
          - \logmsos^2
          \bigg]
\nonumber\\&&\mbox{}
       + \left({M_t^2\over s}\right)^{2} \, \bigg[
          {13\over 4}
          - 11\,\zeta_2
          + {61\over 6}\,\logmsos
          - {3\over 2}\,\logmsos^2
          \bigg]
\nonumber\\&&\mbox{}
       + \left({M_t^2\over s}\right)^{3} \, \bigg[
          {3851\over 162}
          - {160\over 9}\,\zeta_2
          + {563\over 27}\,\logmsos
          - 6\,\logmsos^2
          \bigg]
\nonumber\\&&\mbox{}
       + \left({M_t^2\over s}\right)^{4} \, \bigg[
          {199907\over 2592}
          - {460\over 9}\,\zeta_2
          + {10567\over 216}\,\logmsos
          - {39\over 2}\,\logmsos^2
          \bigg]
\bigg\}
 + \ldots\,\,,
         \\[.4cm]
   R^{(2),p}_{\it F} &=&  3 \, \bigg\{
       - {13\over 12}
          - {3\over 2}\,\zeta_2
          + \zeta_3
          - {13\over 12}\,\logmsos
          + {1\over 4}\,\logmsos^2
\nonumber\\&&\mbox{}
       + {M_t^2\over s} \, \bigg[
          - {11\over 3}
          + 6\,\zeta_2
          - 2\,\zeta_3
          + {13\over 3}\,\logmsos
          - \logmsos^2
          \bigg]
\nonumber\\&&\mbox{}
       + \left({M_t^2\over s}\right)^{2} \, \bigg[
          - {17\over 4}
          + 7\,\zeta_2
          + {35\over 3}\,\logmsos
          - {3\over 2}\,\logmsos^2
          \bigg]
\nonumber\\&&\mbox{}
       + \left({M_t^2\over s}\right)^{3} \, \bigg[
          {155\over 6}
          + {226\over 9}\,\zeta_2
          + {113\over 9}\,\logmsos
          - {31\over 9}\,\logmsos^2
          \bigg]
\nonumber\\&&\mbox{}
       + \left({M_t^2\over s}\right)^{4} \, \bigg[
          {52783\over 864}
          + {977\over 18}\,\zeta_2
          - {823\over 72}\,\logmsos
          + {101\over 36}\,\logmsos^2
          \bigg]
\bigg\}
 + \ldots\,\,,
         \\[.4cm]
   R^{(2),p}_{\it S} &=&  3 \, \bigg\{
         \left({M_t^2\over s}\right)^{2} \, \bigg[
          6
          + 36\,\zeta_3
          \bigg]
       + \left({M_t^2\over s}\right)^{3} \, \bigg[
          {363\over 8}
          - 36\,\zeta_2
          + 18\,\zeta_3
          - 63\,\logmsos
          + 18\,\logmsos^2
          \bigg]
\nonumber\\&&\mbox{}
       + \left({M_t^2\over s}\right)^{4} \, \bigg[
          - {7727\over 162}
          - 132\,\zeta_2
          + 40\,\zeta_3
          - 86\,\logmsos
          + 66\,\logmsos^2
          \bigg]
\bigg\}
 + \ldots\,\,.
\end{eqnarray}

The decay width of a scalar and 
pseudo-scalar Higgs boson is then given by
\begin{eqnarray}
\Gamma(H/A\to t\bar{t}) &=& \frac{G_FM_HM_t^2}{4\sqrt{2}\pi}
  \left[R^{s/p}(M_{H/A}^2) 
 -C_F\,T\,\left(\frac{\alpha_s}{\pi}\right)^2\,R_{gg}^{(2),s/p}(M_{H/A}^2)
  \right],
\label{eqwidth}
\end{eqnarray}
where $R_{gg}^{(2),\delta}$ is the contribution from the pure gluonic
cut appearing in the imaginary part of the singlet diagrams which has to
be subtracted if we are interested only in the fermionic final states.
$R_{gg}^{(2),\delta}$ is known analytically both for the scalar
\cite{EllGaiNan76} and pseudo-scalar case \cite{DjoSpiZer93}.  For
completeness we list the expansion in $M_t^2/s$:
\begin{eqnarray}
R_{gg}^{(2),s}(s) &=& 3\Bigg\{
\frac{M_t^2}{s}\left[
2
+ 6\zeta_2
+ \frac{45}{4}\zeta_4
+ \left(-1 + \frac{3}{2}\zeta_2\right) \logmsos^2
+ \frac{1}{8}\logmsos^4 
\right]
\nonumber\\
&&
+ \left(\frac{M_t^2}{s}\right)^2\left[
- 24\zeta_2
- 90\zeta_4
+ \left(-4 + 6\zeta_2\right) \logmsos
+ \left(4 - 12\zeta_2\right) \logmsos^2
\right.\nonumber\\&&\mbox{\hspace{.5cm}}\left.
+ \logmsos^3
- \logmsos^4 
\right]
\nonumber\\
&&
+ \left(\frac{M_t^2}{s}\right)^3
\left[
- 4
+ 6\zeta_2
+ 180\zeta_4
+ \left(10 - 39\zeta_2\right) \logmsos
+ \left(3 + 24\zeta_2\right) \logmsos^2
\right.\nonumber\\&&\mbox{\hspace{.5cm}}\left.
- \frac{13}{2}\logmsos^3
+ 2\logmsos^4 
\right]
\nonumber\\
&&
+  \left(\frac{M_t^2}{s}\right)^4\left[
  4
- 30\zeta_2
+ \left(\frac{44}{3} + 44\zeta_2\right) \logmsos
- 15\logmsos^2 
+ \frac{22}{3}\logmsos^3
\right]
\Bigg\}
 + \ldots \,\,,
\nonumber\\
\\
R_{gg}^{(2),p}(s) &=& 3\Bigg\{
\frac{M_t^2}{s}\left[
 \frac{45}{4}\zeta_4
+ \frac{3}{2}\zeta_2\logmsos^2
+ \frac{1}{8}\logmsos^4
\right]
+\left(\frac{M_t^2}{s}\right)^2\left(
  6\zeta_2\logmsos
+ \logmsos^3
\right)
\nonumber\\
&&
+ \left(\frac{M_t^2}{s}\right)^3\left[
 6\zeta_2
+ 9\zeta_2\logmsos
+ 3 \logmsos^2
+ \frac{3}{2}\logmsos^3
\right]
\nonumber\\
&&
+ \left(\frac{M_t^2}{s}\right)^4\left[
  18\zeta_2
+ \left(4 + 20\zeta_2\right)\logmsos
+ 9\logmsos^2
+ \frac{10}{3}\logmsos^3
\right]
\Bigg\}
 + \ldots\,\,.
\end{eqnarray}
Note that the analytical structure is not changed 
if $M_t$ is replaced by the $\overline{\mbox{MS}}$ 
mass because $R_{gg}^{(2),\delta}$
only contributes at order $\alpha_s^2$.
For the numerical results presented below we only use the inclusive
quantity $R^\delta(s)$.

The results for the constant and first mass corrections are in agreement
with the literature both for the scalar
\cite{Sur94,CheKwi96}
and pseudo-scalar case
\cite{Sur942}.
A strong check is provided for the terms proportional to $n_l$
where a successful comparison with the exact results for the
scalar
\cite{Mel96,HoaTeu97}
and pseudo-scalar 
\cite{HoaTeu97}
contributions was 
possible\footnote{We would like to thank the authors of 
\cite{HoaTeu97} for providing us with the results prior to
publication.}.

\begin{figure}[t]
 \begin{center}
 \begin{tabular}{cc}
   \leavevmode
   \epsfxsize=6.5cm
   \epsffile[110 270 480 560]{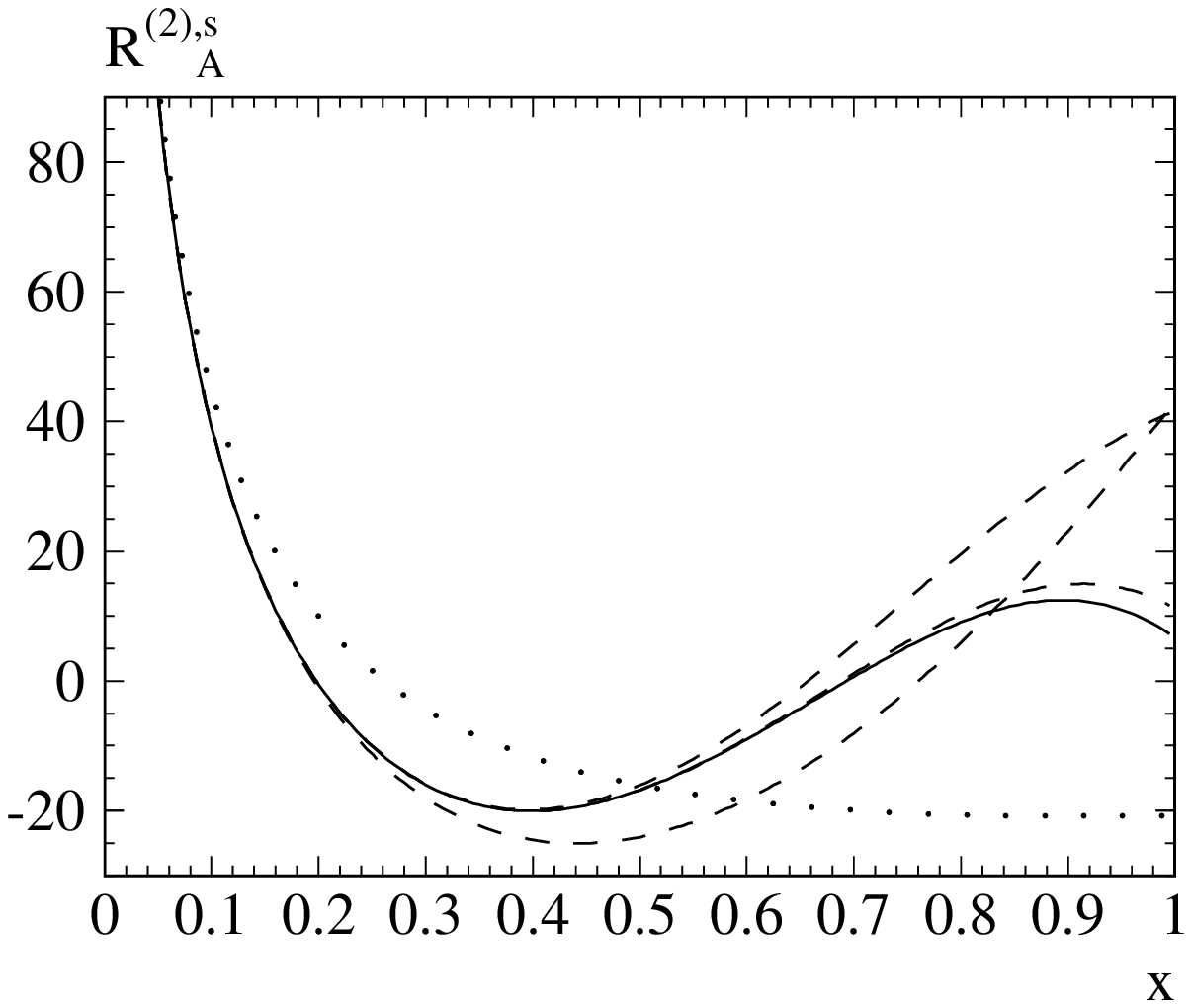 }
   &
   \epsfxsize=6.5cm
   \epsffile[110 270 480 560]{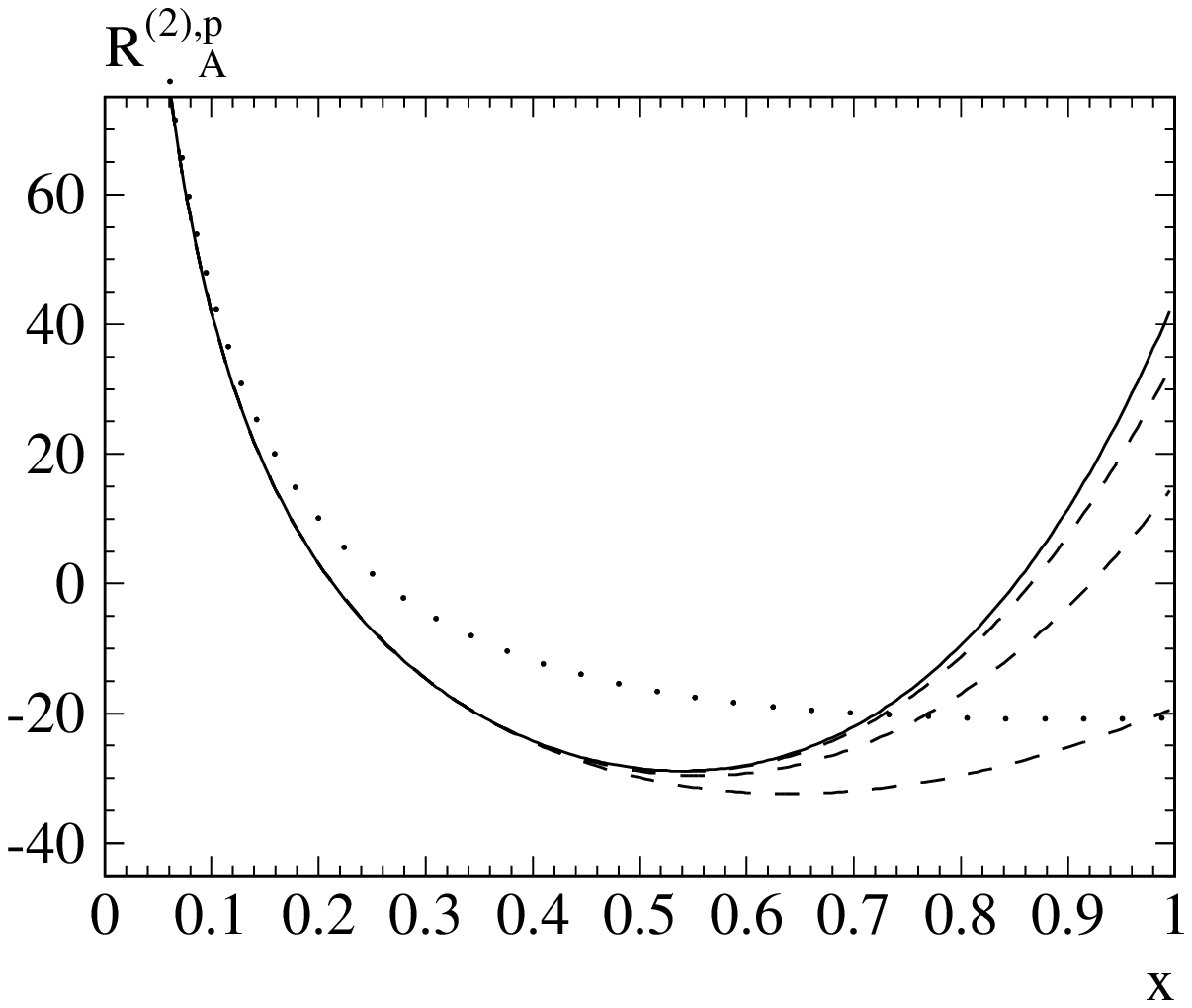 }
\\
   \epsfxsize=6.5cm
   \epsffile[110 270 480 560]{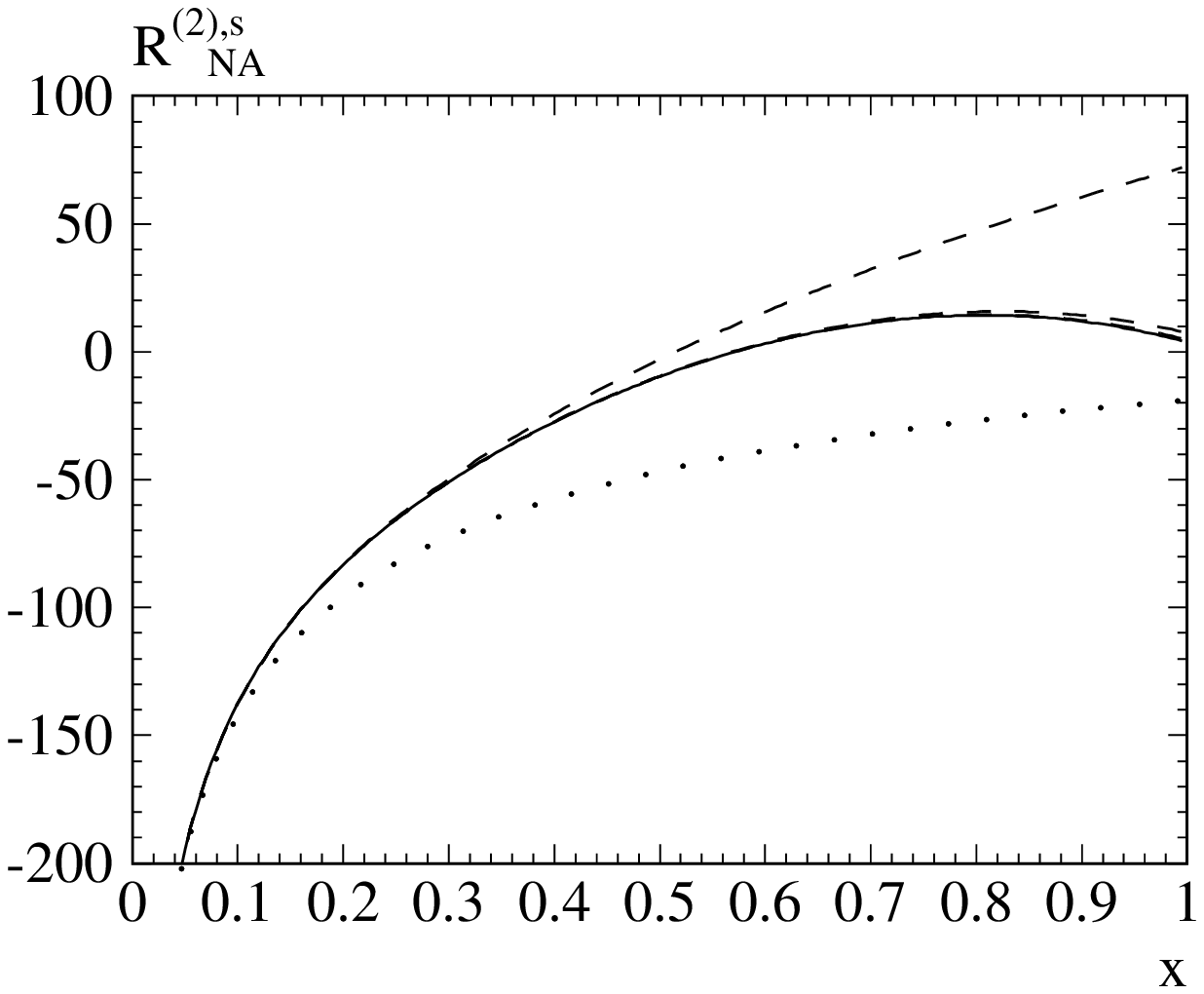 }
   &
   \epsfxsize=6.5cm
   \epsffile[110 270 480 560]{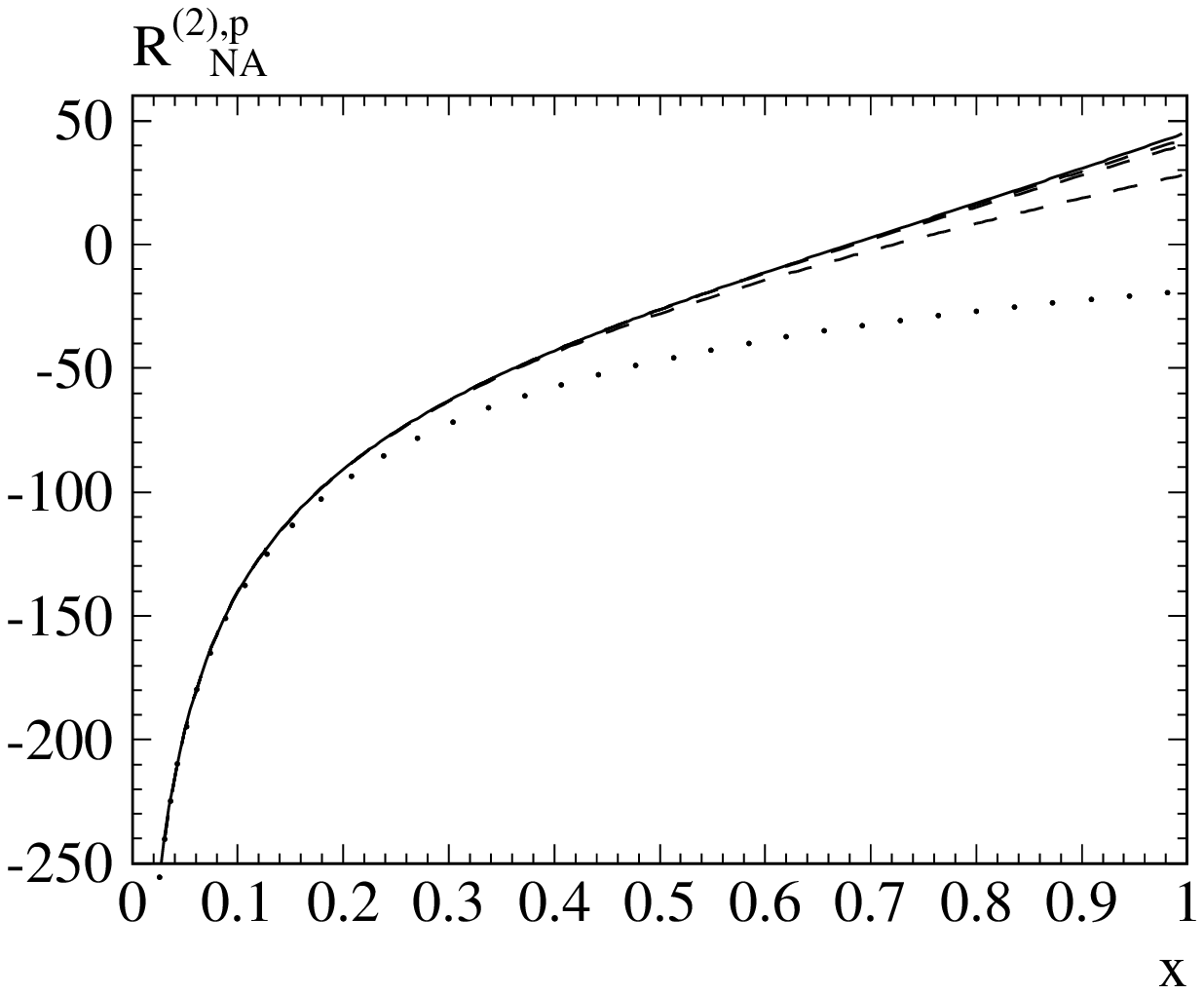 }
\\
   \epsfxsize=6.5cm
   \epsffile[110 270 480 560]{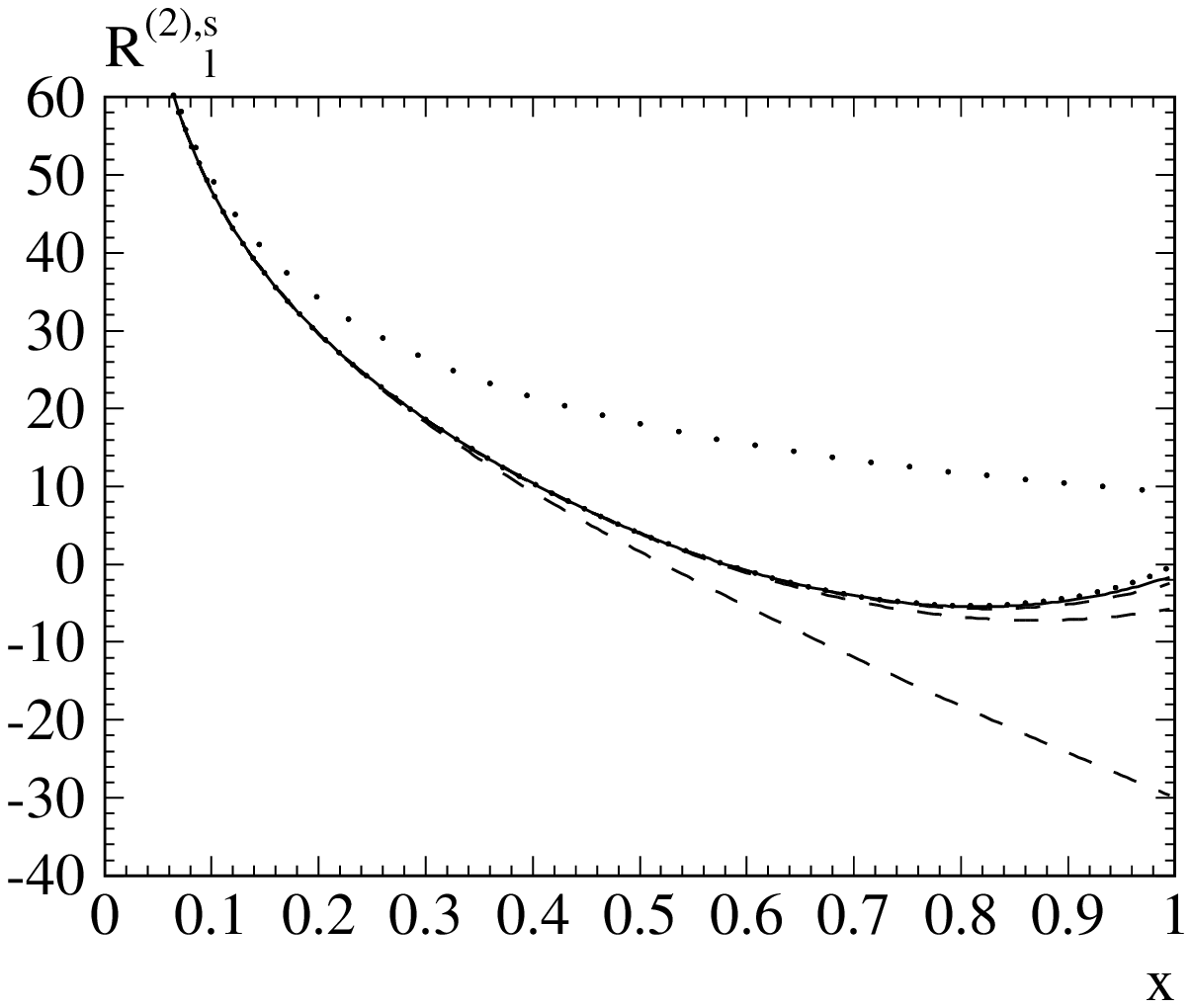 }
   &
   \epsfxsize=6.5cm
   \epsffile[110 270 480 560]{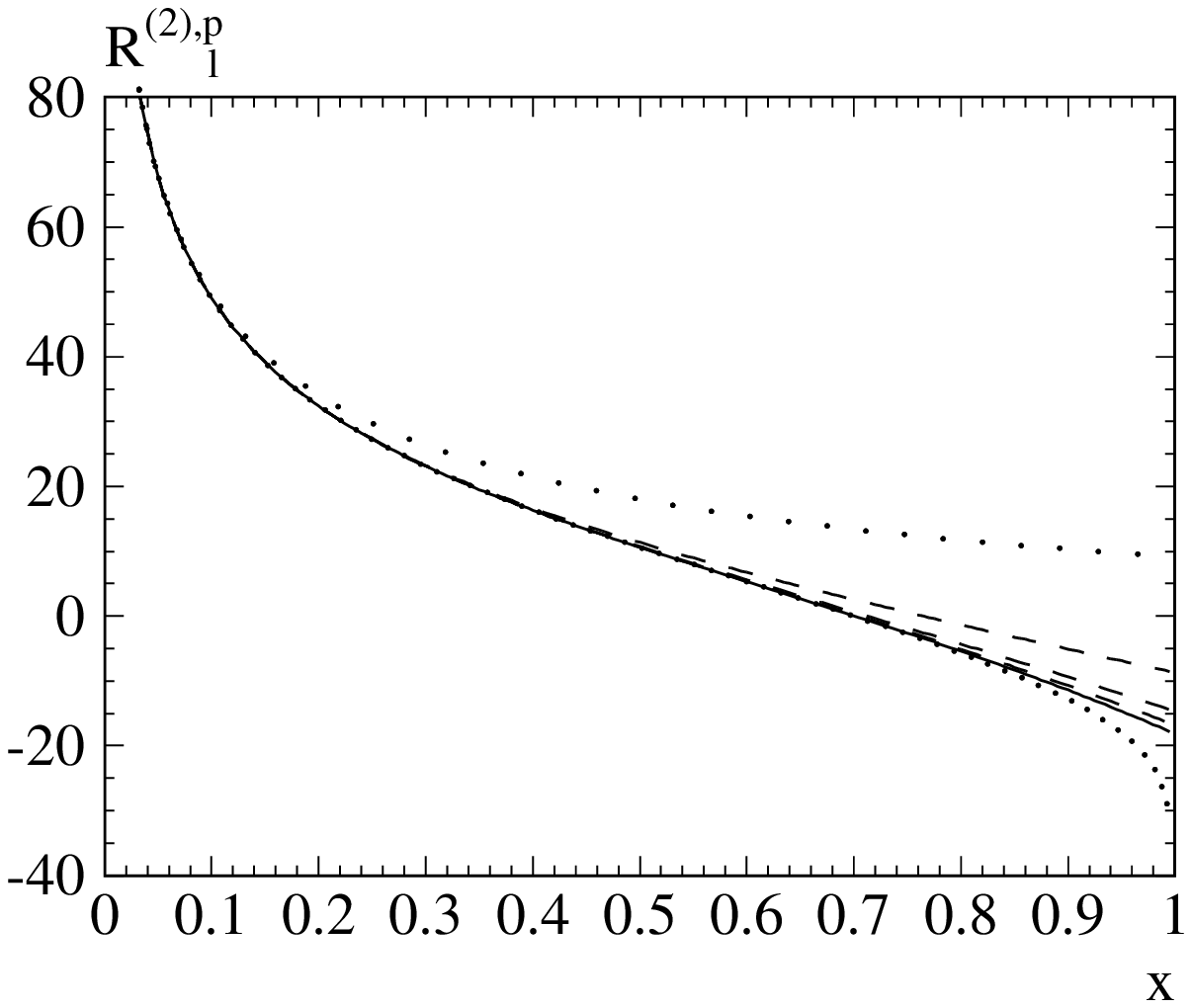 }
 \end{tabular}
 \caption{\label{figspanal} $R^{(2),\delta}_n$ with $\delta=s,p$ and 
                            $n=A,NA,l$ plotted against $x$. 
                            The same notation as in Fig.~\ref{figsp01}  
                            is adopted.
}
 \end{center}
\end{figure}

\begin{figure}[ht]
 \begin{center}
 \begin{tabular}{cc}
   \leavevmode
   \epsfxsize=6.5cm
   \epsffile[110 270 480 560]{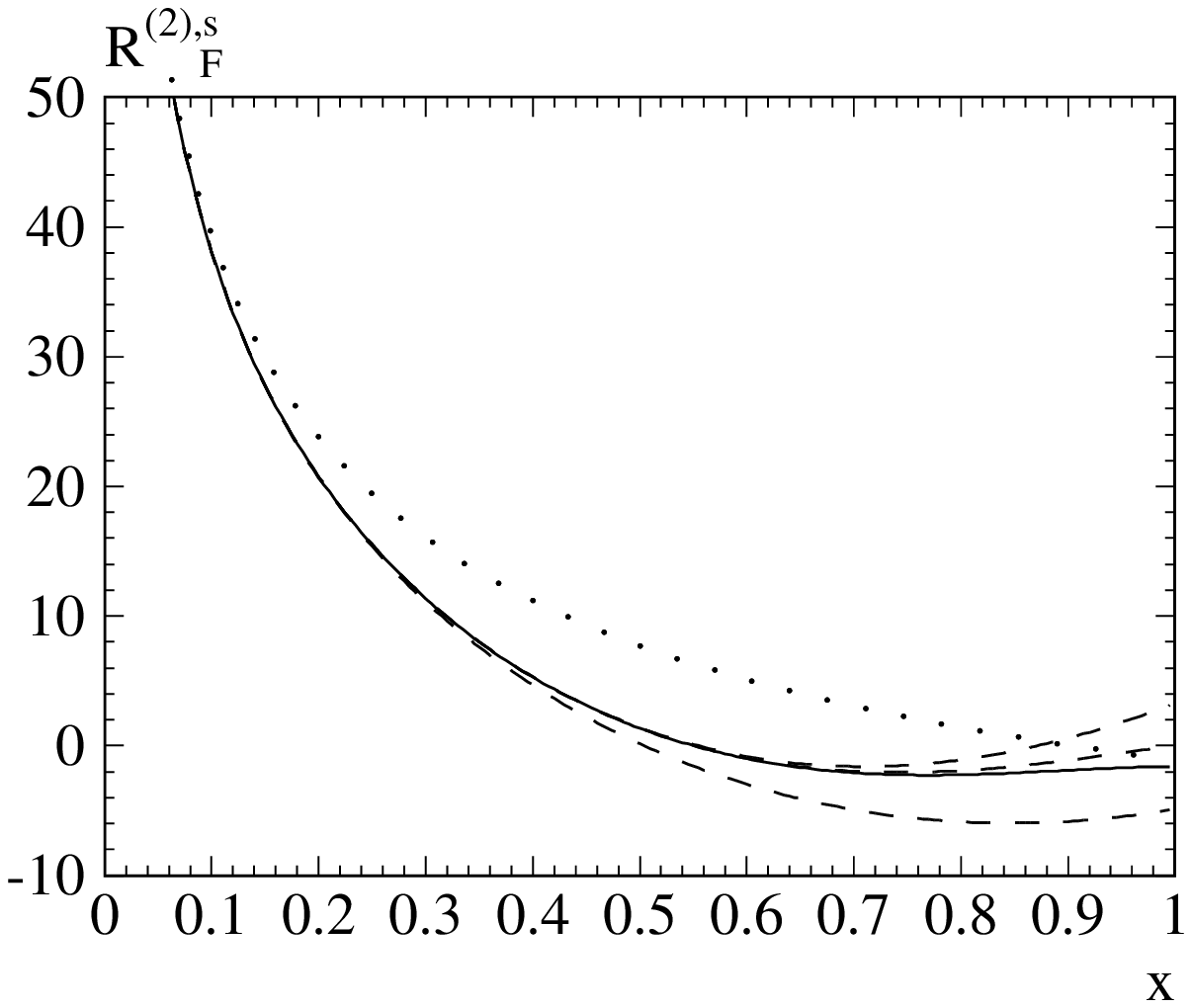 }
   &
   \epsfxsize=6.5cm
   \epsffile[110 270 480 560]{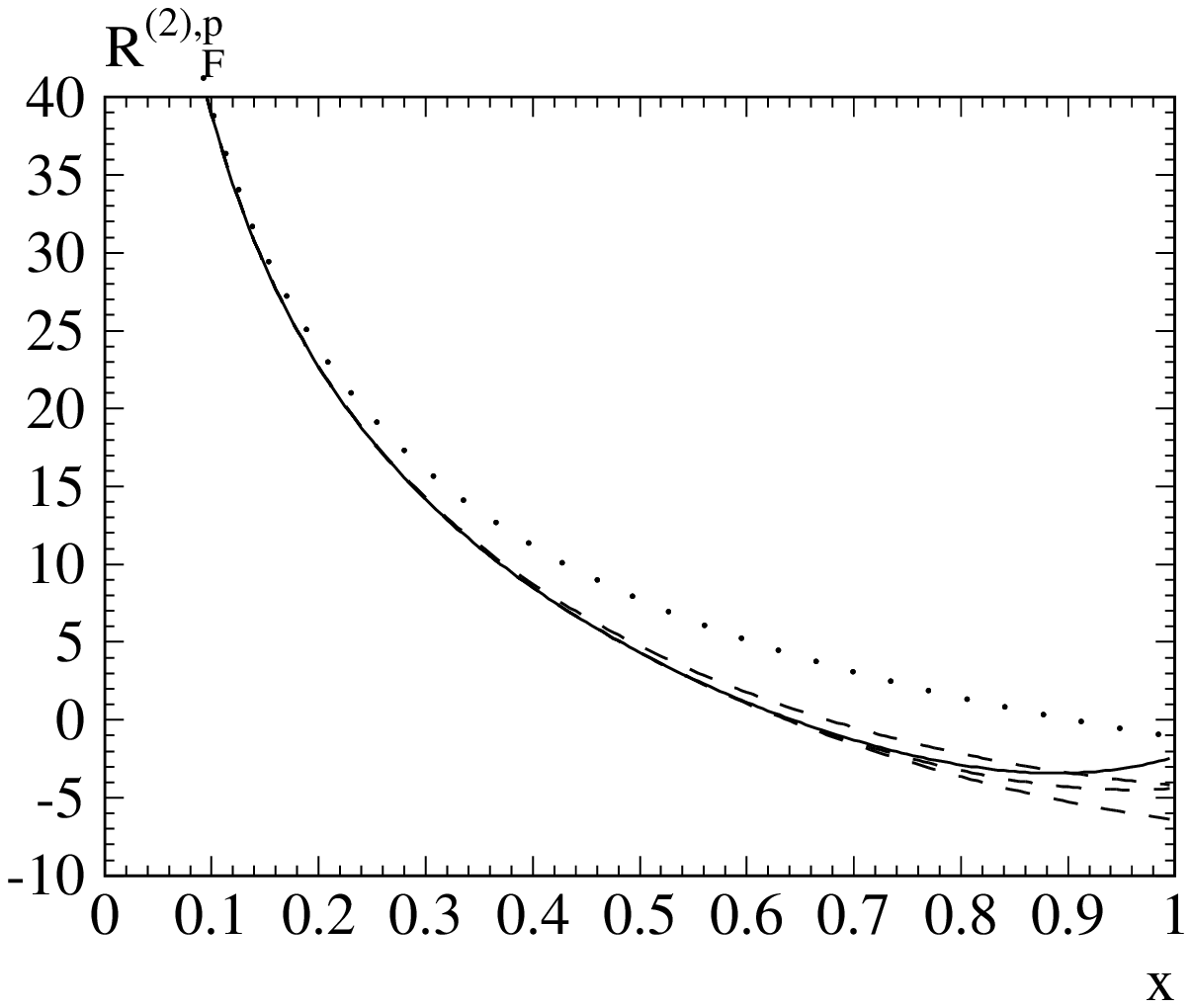 }
\\
   \epsfxsize=6.5cm
   \epsffile[110 270 480 560]{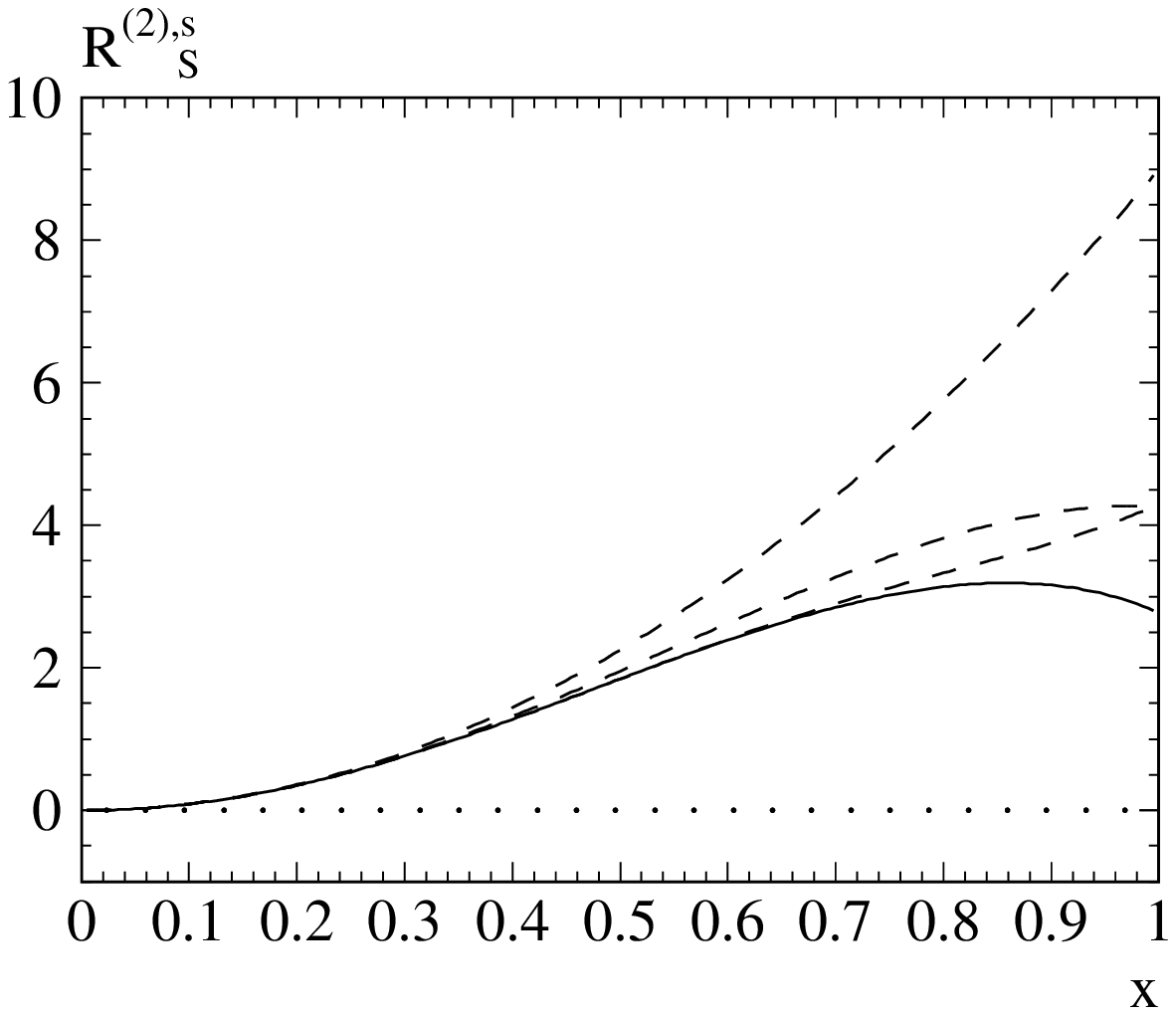 }
   &
   \epsfxsize=6.5cm
   \epsffile[110 270 480 560]{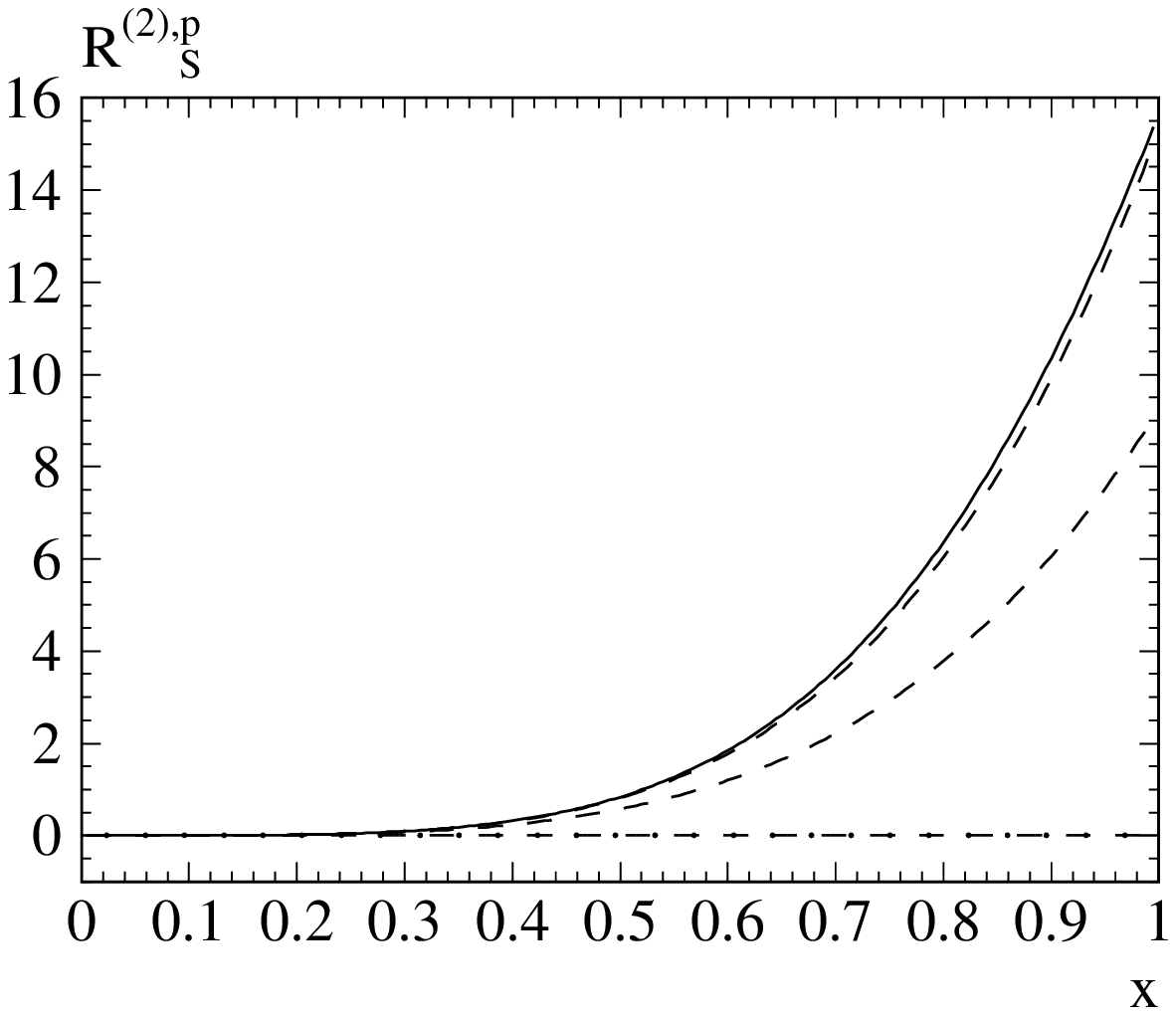 }
 \end{tabular}
 \caption{\label{figspfs} $R^{(2),\delta}_n$ with $\delta=s,p$ and 
                            $n=F,S$ plotted against $x$. 
                            The same notation as in Fig.~\ref{figsp01}  
                            is adopted.
}
 \end{center}
\end{figure}

In Figs.~\ref{figspanal} and \ref{figspfs} the contributions of the
different colour factors are plotted including successively higher
orders in $M_t^2/s$.  The notation is the same as in Fig.~\ref{figsp01}.
Motivated by the behavior of the Born- and one-loop terms of
Fig.~\ref{figsp01}, one may consider as a measure of validity 
for each curve 
the range in $x$ where it coincides with the one containing most
powers of $x$ (i.e., the solid line in Figs.~\ref{figspanal} and
\ref{figspfs}).  
This criterion is further
justified by the light fermion contribution $R_{\it l}^{(2),\delta}$,
where also the exact result is known.
For the scalar case, $R_{\it l}^{(2),s}$, the agreement 
of the exact result and the $M_t^8/s^4$-terms
is almost perfect up to $x=1$, while for $R_{\it l}^{(2),p}$ the
convergence above $x=0.9$ gets poorer, but the above argument concerning
the validity holds perfectly in both cases.  

However, one objection is in order here, namely the fact that 
$R_{\it A}^{(2),\delta}$, $R_{\it NA}^{(2),\delta}$, $R_{\it F}^{(2),\delta}$
and $R_{\it S}^{(2),\delta}$, in contrast to $R_{\it l}^{(2),\delta}$,
exhibit a four-particle threshold at $x=1/2$ which may spoil convergence
for $x > 1/2$. Indeed, the curves for $R_{\it F}^{(2),\delta}$, 
$R_{\it S}^{(2),\delta}$ (Fig.~\ref{figspfs}) develop a relatively large
spread in this $x$-range. Nevertheless, the range of validity as defined
above extends to $x=0.7-0.8$.  Despite this objection, the
curves for $R_{\it NA}^{(2),\delta}$ seem to converge very well again
almost up to $x=1$, those for $R_{\it A}^{(2),\delta}$ at least up to 
$x\approx 0.85$. In particular for the scalar case, where the exact result
is known to be zero\footnote{$R_S^{(2),s}\not=0$ because the gluon cut
is still present.}
at $x=1$, the behavior of the approximations is
quite promising.

Summarizing, one can see that while the quadratic mass terms seem to
reasonably approximate the result up to $x=0.3-0.4$ for the scalar and
$x\approx 0.5$ for the pseudo-scalar case, the quartic terms
considerably improve the expansions in most cases up to values 
of $x\approx 0.85$, in some cases even up to $x=1$. The influence of the 
higher order terms is rather small and practically only needed for
$R_A^{(2),\delta}$ and $R_S^{(2),\delta}$.

\begin{table}[th]
{\footnotesize
\renewcommand{\arraystretch}{1.4}
\begin{center}
\begin{tabular}{|l||r|r|r|r|r||l||l|}
 \hline 
   \mbox{} & ${(M_t^2)^0}$ &  ${(M_t^2)^1}$ & 
      ${(M_t^2)^2}$ &  ${(M_t^2)^3}$ & 
      ${(M_t^2)^4}$ &  $\Sigma$   &  exact \\ 
 \hline \hline
 \mbox{} & \multicolumn{5}{|c||}{scalar, on-shell} & &  \\ 
  \hline 
$R^{(0),s}/3$ & $
               1.000$ & $
              -0.907$ & $
               0.137$ & $
               0.014$ & $
               0.003$ & $
       0.247$ & $       0.248$ \\ 
 \hline 
$C_FR^{(1),s}/3$ & $
              -0.778$ & $
               5.646$ & $
              -3.080$ & $
               0.010$ & $
               0.004$ & $
       1.802$ & $       1.802$ \\ 
 \hline 
$R^{(2),s}/3$ & $
             -35.803$ & $
              77.056$ & $
             -17.792$ & $
              -5.347$ & $
              -0.680$ & $
      17.435$ & $-$ \\ 
 \hline 
 $\Sigma_i (\alpha_s/\pi)^i$ 
&$  0.943$
&$ -0.663$
&$  0.026$
&$  0.009$
&$  0.003$
&$  0.318$ & \mbox{}\\ \hline \hline
 \mbox{} & \multicolumn{5}{|c||}{scalar, $\overline{\rm MS}$} & &  \\ 
  \hline 
$\frac{m_t^2}{M_t^2}\bar{R}^{(0),s}/3$ & $
 0.784$
&$ -0.558$
&$  0.066$
&$  0.005$
&$  0.001$
&$  0.298$
&$  0.299  $ 
\\
 \hline 
$\frac{m_t^2}{M_t^2}C_F\bar{R}^{(1),s}/3$ & $
  4.445$
&$ -3.723$
&$ -0.238$
&$  0.143$
&$  0.032$
&$  0.659$
&$  0.673  $
\\
 \hline 
$\frac{m_t^2}{M_t^2}\bar{R}^{(2),s}/3$ & $
 21.799$
&$ -7.606$
&$-15.334$
&$  0.680$
&$  0.546$
&$  0.086$
&$  -  $
\\
 \hline 
 $\Sigma_i (\alpha_s/\pi)^i$ 
&$  0.941$
&$ -0.679$
&$  0.045$
&$  0.010$
&$  0.002$
&$  0.319$
& \mbox{}\\ \hline \hline
 \mbox{} & \multicolumn{5}{|c||}{pseudo-scalar, on-shell} & &  \\ 
  \hline 
$R^{(0),p}/3$ & $
               1.000$ & $
              -0.302$ & $
              -0.046$ & $
              -0.014$ & $
              -0.005$ & $
       0.633$ & $       0.629$ \\ 
 \hline 
$C_FR^{(1),p}/3$ & $
              -0.778$ & $
               3.495$ & $
               0.417$ & $
               0.047$ & $
               0.027$ & $
       3.208$ & $       3.238$ \\ 
 \hline 
$R^{(2),p}/3$ & $
             -35.803$ & $
              25.024$ & $
              12.173$ & $
               3.780$ & $
               1.293$ & $
       6.467$ & $-$ \\ 
 \hline 
 $\Sigma_i (\alpha_s/\pi)^i$ 
&$  0.943$
&$ -0.172$
&$ -0.022$
&$ -0.009$
&$ -0.003$
&$  0.737$ & \mbox{}\\ \hline \hline
 \mbox{} & \multicolumn{5}{|c||}{pseudo-scalar, $\overline{\rm MS}$} & &  \\ 
  \hline 
$\frac{m_t^2}{M_t^2}\bar{R}^{(0),p}/3$ & $
  0.784$
&$ -0.186$
&$ -0.022$
&$ -0.005$
&$ -0.002$
&$  0.569$
&$  0.569  $
\\
 \hline 
$\frac{m_t^2}{M_t^2}C_F\bar{R}^{(1),p}/3$ & $
  4.445$
&$ -0.248$
&$ -0.215$
&$ -0.119$
&$ -0.043$
&$  3.821$
&$  3.791  $
\\
 \hline 
$\frac{m_t^2}{M_t^2}\bar{R}^{(2),p}/3$ & $
 21.799$
&$  9.854$
&$  2.455$
&$ -0.781$
&$ -0.520$
&$ 32.807$
&$ -  $
\\
 \hline 
 $\Sigma_i (\alpha_s/\pi)^i$
&$  0.941$
&$ -0.185$
&$ -0.026$
&$ -0.010$
&$ -0.003$
&$  0.717$
&\mbox{}\\\hline 
\end{tabular}
\end{center}
}
\caption{\label{tabsp}
  Numerical results for $R^s$ and $R^p$ both in the on-shell and
  $\overline{\mbox{MS}}$ scheme.  The contributions from the mass terms
  $(M_t^2)^i$, their sum ($\Sigma$) and, where available, the exact
  results are shown.  $\Sigma_i(\alpha_s/\pi)^i$ is the sum of the 1-,
  2- and 3-loop terms.  The numbers correspond to $M_t=175$~GeV and
  $M_{H/A}=450$~GeV.  The renormalization scale $\mu^2$ is set to
  $s=M_{H/A}^2$.  }
\end{table}

It is interesting to compare the results both for the on-shell and the
$\overline {\rm MS}$-definition of the quark mass.
Therefore we set $\mu^2=s$ which is the natural choice because it
eliminates all logarithms of the constant and quadratic terms in the
$\overline{\mbox{MS}}$ scheme.  Starting from the quartic term, however,
$\ln M_t^2/s$ terms remain, resulting from non-trivial operators which
are absent before \cite{CheSpi87}.  In contrast to the
$\overline{\mbox{MS}}$ scheme the results in the on-shell scheme develop
$\ln M_t^2/s$ terms starting from the lowest order.  
It is nevertheless
instructive to compare the expansions both in $M_t^2/s$ and
$\alpha_s(s)$ in the two schemes.  Therefore 
the renormalization group invariant quantities 
$m_t^2(s)\bar{R}^\delta(s)/M_t^2$ and $R^\delta(s)$ are considered.
We choose $n_l=5$,
$\alpha_s^{(5)}(M_Z^2)=0.118$, $M_t=175$~GeV and $M_{H/A}=450$~GeV
which corresponds to $x\approx0.78$.  The corresponding
$\overline{\mbox{MS}}$ top mass, $m_t(M_{H/A})$, evaluates to
$m_t(450~\mbox{GeV})=155$~GeV and the two-loop beta function leads to
$\alpha_s((450~\mbox{GeV})^2)\equiv
\alpha_s^{(6)}((450~\mbox{GeV})^2)\approx0.096$.  In Tab.~\ref{tabsp}
the numbers are listed.  It can be seen that the quartic terms are still
sizeable and in many cases numerically of the same order as the constant
and quadratic ones.  They become particularly important when the first
two terms happen to cancel to a large extent.  The higher
order terms give only small contributions at least in the three-loop
case.

Usually in the $\overline{\mbox{MS}}$ scheme a faster
convergence of the QCD series is expected in the sense that 
the coefficients in front of $(\alpha_s/\pi)^n$ are smaller than 
in the on-shell scheme. 
Considering the sum of all approximation terms available
this happens to be true only for the scalar correlator
whereas in the pseudo-scalar case it is vice versa. 
A possible explanation might be that the coefficient of the leading term
in the on-shell scheme is already rather small
because of an accidental
cancellation between the constant and the logarithm.
For the scalar correlator there is in addition a large
cancellation between the first three terms
in the $\overline{\mbox{MS}}$ scheme
so that the sum turns out to be also rather small.
This is not the case for $R^p$ where the first three terms
have the same sign.
However, the scheme dependence is drastically reduced by the
inclusion of the ${\cal O}(\alpha_s^2)$ terms as can be seen 
in Tab.~\ref{tabsp}.

Finally we want to present numerical values for the physical decay rate
into top quarks. For convenience the square of the on-shell mass,
$M_t^2$, is factored out both in the on-shell and $\overline{\mbox{MS}}$
scheme.  Using Eq.~(\ref{eqwidth}) the expansions of the decay width for
the scalar and pseudo-scalar case look like:
\begin{eqnarray}
\Gamma(H\to t\bar{t}) &=& \frac{G_FM_HM_t^2}{4\sqrt{2}\pi}
\left[
0.7404 + 0.1652 + 0.0469  + 0.0019 - 0.0022  
\right],
\\
\bar{\Gamma}(H\to t\bar{t}) &=& \frac{G_FM_HM_t^2}{4\sqrt{2}\pi}
\left[
0.8954 + 0.0604 - 0.0011  + 0.0014 - 0.0020 
\right],
\\
\Gamma(A\to t\bar{t}) &=& \frac{G_FM_HM_t^2}{4\sqrt{2}\pi}
\left[ 
1.8982 + 0.2941 + 0.0146 + 0.0035 - 0.0051 
\right],
\\
\bar{\Gamma}(A\to t\bar{t}) &=& \frac{G_FM_HM_t^2}{4\sqrt{2}\pi}
\left[
1.7084 + 0.3503 + 0.0903  + 0.0016 - 0.0038 
\right],
\end{eqnarray}
where the first and second number correspond to the Born and
${\cal O}(\alpha_s)$ correction. The ${\cal O}(\alpha_s^2)$ 
terms are separated into the non-singlet (third number) and 
singlet (fourth number) contribution. The last number corresponds
to the gluonic cut which has to be subtracted. 
The last two numbers cancel each other to a large extent
which means that the two-gluon-cut dominates the 
imaginary part of the double-triangle diagram.
Furthermore one observes that the two-loop QCD corrections
amount up to $\approx5\%$.

To conclude, we have computed analytically
the first five terms of the scalar and
pseudo-scalar current correlator in the expansion 
for large external momentum $q$.
As an application the decay of a scalar and pseudo-scalar 
Higgs boson into top quarks is considered.
In this case the higher order mass corrections, especially the
quartic terms, emerge to be important.

\centerline{\bf Acknowledgments}
\smallskip\noindent
We would like to thank K.G. Chetyrkin and J.H. K\"uhn 
for fruitful discussions and for carefully reading the manuscript.
M.S. would like to thank P. Gambino for useful discussions.
The work of R.H. was supported by the ``Landesgraduiertenf\"orderung'' 
at the University of Karlsruhe.


\end{document}